\documentclass[aps,prx,twocolumn,superscriptaddress,amsmath,nofootinbib,amssymb,10pt,floatfix]{revtex4-1}
\usepackage{graphicx}
\usepackage{mathtools}
\usepackage{ulem}   
\usepackage[usenames,dvipsnames]{xcolor}
\usepackage{bbold}
\usepackage{comment}
\usepackage[colorlinks, linkcolor=blue, citecolor=blue]{hyperref}

\renewcommand{\phi}{\varphi}
\renewcommand{\epsilon}{\varepsilon}
\def \td{\mathrm{d}}
\newcommand{\defeq}{\vcentcolon=}
\renewcommand{\vec}[1]{\boldsymbol{#1}}

\renewcommand{\sc}[1]{\mathcal{#1}}
\newcommand{\bb}[1]{\mathbb{#1}}

\begin{document}
\title{Metal to Wigner-Mott insulator transition in  two-leg ladders}

\author{Seth Musser}
\author{T. Senthil}
\affiliation{Department of Physics, Massachusetts Institute of Technology, Cambridge, Massachusetts 02139, USA}
\begin{abstract}
We study theoretically the quantum phase transition from a metal to a Wigner-Mott insulator at fractional commensurate filling on a two-leg ladder. We show that a continuous transition out of a symmetry-preserving Luttinger liquid metal is possible where the onset of insulating behavior is accompanied by the breaking of the lattice translation symmetry. At fillings $\nu = 1/m$ per spin per unit cell, we find that the spin degrees of freedom also acquire a gap at the Wigner-Mott transition for odd integer $m$. In contrast for even integer $m$, the spin sector remains gapless and the resulting insulator is a ladder analog of the two-dimensional spinon surface state. In both cases, a charge neutral spinless mode remains gapless across the Wigner-Mott transition. We discuss physical properties of these transitions, and comment on insights obtained for thinking about continuous Wigner-Mott transitions in two-dimensional systems which are being studied in moire materials. 
\end{abstract}

\maketitle

\section{Introduction}
Despite decades of study, the vicinity of the Mott metal-insulator transition continues to surprise and challenge theoretical physics.  Particularly fascinating is the possibility of a continuous second-order Mott transition across which the metallic Fermi surface (FS) must disappear while maintaining its size \cite{senthil_critical_2008}. Recent experiments \cite{li_continuous_2021, ghiotto_quantum_2021} on moire superlatices formed from transition metal dichalcogenide (TMD) materials have found good evidence for such a continuous Mott transition at half filling of a band. A theory \cite{senthil_theory_2008} for a continuous Mott transition from a Fermi liquid metal to a symmetry-preserving Mott insulator in a certain quantum spin liquid phase exists. A refinement of this theory to include effects of various kinds of disorder \cite{kim_continuous_2022, unpublishedSenthil} seems to be able to account for the observed electrical transport in Refs. \cite{li_continuous_2021, ghiotto_quantum_2021}.

\begin{figure}
    \centering
    \includegraphics[width=\columnwidth]{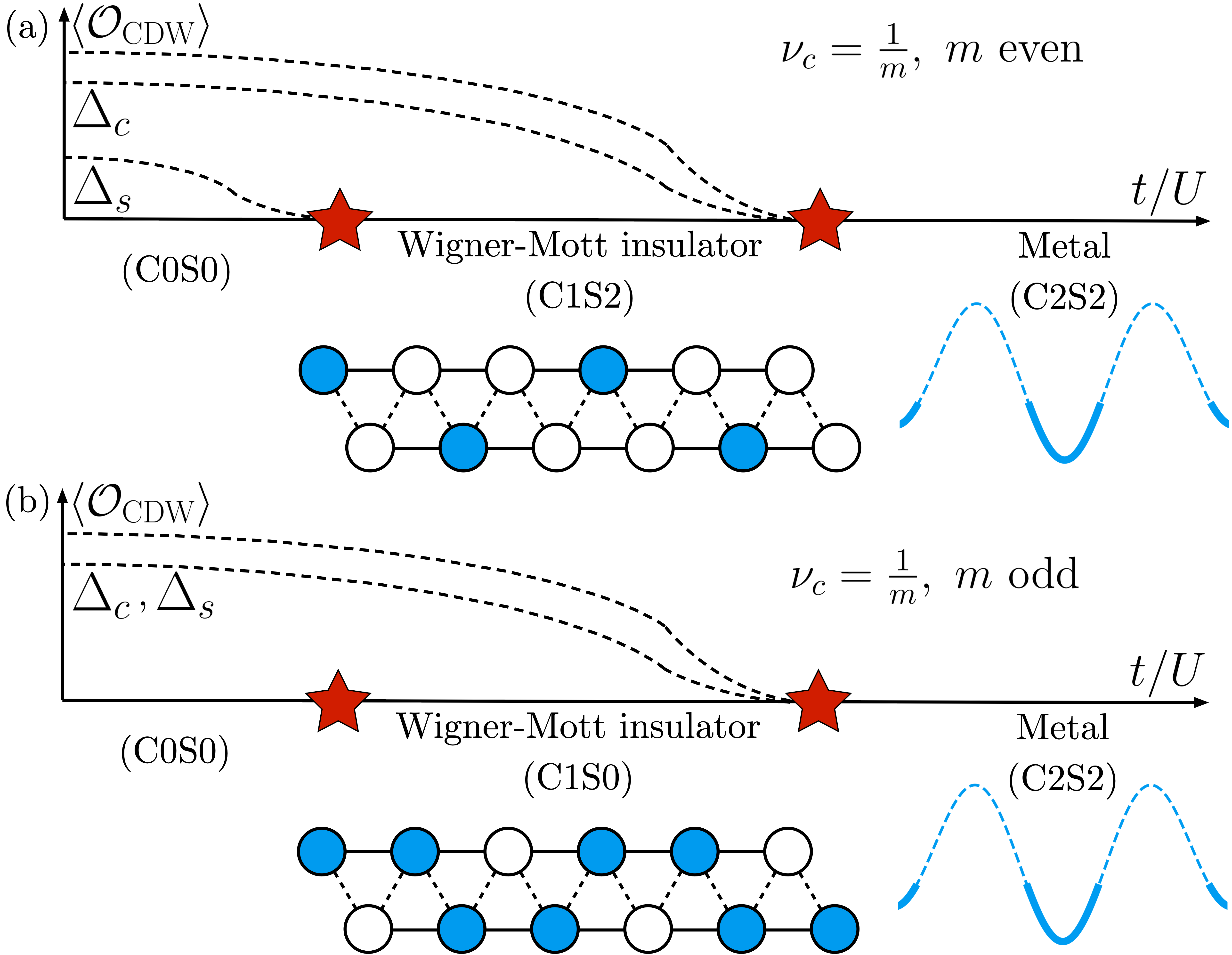}
    \caption{A continuous metal to Wigner-Mott insulating transition in the triangular strip at fillings per spin of $\nu_c = 1/m$ with: (a) $m$ even, and (b) $m$ odd. On the metallic side, the system exhibits a dispersion with well-defined Fermi points separating filled from unfilled states. This dispersion is shown below the metallic phase in both (a) and (b). Interactions, labeled here by $U/t$, are then increased while keeping the filling fixed. The system eventually crosses a critical point, shown as the rightmost red star in both (a) and (b), where it develops both: a charge gap, $\Delta_c$, and a nonzero expectation value for translation symmetry breaking order parameters, $\langle \mathcal{O}_{\mathrm{CDW}} \rangle \neq 0$. This translation symmetry broken state is labeled the Wigner-Mott insulating state and a possible real space charge arrangement is shown for $\nu_c = 1/6$ in (a) and $\nu_c = 1/3$ in (b), where the filled blue circles indicate electronic charge. Additionally, for $m$ odd the first critical point will open a spin gap. This is shown in (b). Directly on the insulating side, both (a) and (b) will generically still have a gapless neutral charge mode which will lead to fluctuations of observables at incommensurate wave vectors. This neutral charge mode will be gapped out at a second transition, shown as the rightmost red star in both diagrams. For $m$ even, the transition that gaps out this charge mode will also open a gap in the spin sector. The transition to the fully gapped (C0S0) phase is briefly discussed in Sec. \ref{sec:dominant_umklapp}.}
    \label{fig:WM_defn}
\end{figure}

The moire-TMD setting also enables study of the transition between the metal and a Wigner-Mott (WM) insulator at discrete fractional band fillings. Remarkably, some of these WM transitions also appear to be continuous \cite{tang_dielectric_2022, MakShan}. Theoretically, the possibility of such a continuous (as opposed to one that is first order) Wigner-Mott transition raises a number of fundamental questions which are only beginning to be addressed. For a number of scenarios for how such a transition may proceed, see, e.g., Refs.~\cite{xu_interaction-driven_2022, musser_theory_2022}. Most interesting is the possibility that the charge ordering associated with the WM state develops only in the insulating phase. 
Building on the theory of the continuous Mott transition at half filling, the authors of this paper (together with D. Chowdhury) recently demonstrated the possibility of such a continuous metal to WM insulator transition in two dimensions \cite{musser_theory_2022}. At such a transition, the entire electronic FS disappears abruptly upon approaching from the metallic side, and the insulating charge gap and various order-parameters associated with spontaneously broken space-group symmetries vanish continuously upon approaching from the insulating side. An illustration of this is shown as the rightmost critical point in Fig.~\ref{fig:WM_defn}. As in the figure, transitions of this type might be expected to generically occur for fractional electron fillings of the form $\nu_c = 1/m$, where $m$ is an integer, as any real space arrangement of localized electrons will necessarily break some space-group symmetries.

This previous work provides a proof of principle of the  possibility of such a continuous WM transition in two dimensions. A crucial step was the identification of a low-energy effective field theory that captures both the Fermi liquid and the WM insulator, and thus is capable of describing a direct transition between them. Unfortunately it is hard to control the properties of this theory at the transition point, and Ref. \cite{musser_theory_2022} relied on a 
certain large-N generalization to make calculations tractable. Clearly it will be extremely helpful to study continuous WM transitions in other situations, and see what we can learn.

In contrast to the situation in $d = 2$ space dimensions, the universal properties of the WM transition in a $d = 1$ chain can be understood in an exact way using bosonization methods \cite{schulz_metal-insulator_1994, giamarchi_quantum_2003}. Here we will generalize this treatment to the two-leg triangular ladder (pictured in Fig.~\ref{fig:WM_defn}) as a step toward the full two-dimensional problem.  We will demonstrate that it is indeed possible to have a continuous bandwidth-tuned metal to WM transition in this setup. We will further show that for electron fillings $\nu_c = 1/m$ with $m$ odd this transition must also open a spin gap at the same critical point where the charge is gapped out, shown as the rightmost red star in Fig.~\ref{fig:WM_defn}(b). For even $m$, the spin sector remains gapless across the metal-insulator transition. The resulting insulator is a ladder analog of the two-dimensional spinon FS state in a WM insulator of the kind discussed in Ref. \cite{musser_theory_2022}. For  both even and odd $m$, in the immediate vicinity of the transition, the insulating phase has a gapless spin singlet, charge-neutral excitation mode. This is a new feature of the two-leg ladder not present in the strictly one-dimensional chain and is shown as the intermediate phase in Figs.~\ref{fig:WM_defn}(a) and \ref{fig:WM_defn}(b).

The continuous Mott transition at half filling in a two-leg ladder between a Luttinger liquid metal and a symmetry-preserving gapless Mott insulator was previously studied in Ref. \cite{mishmash_continuous_2015}, and we will build on their analysis. There are a few new features introduced by the extension to the WM transition that we will discuss at various points in the paper.

\section{Introduction to the model and bosonization}

\begin{figure}
    \centering
    \includegraphics[width=\columnwidth]{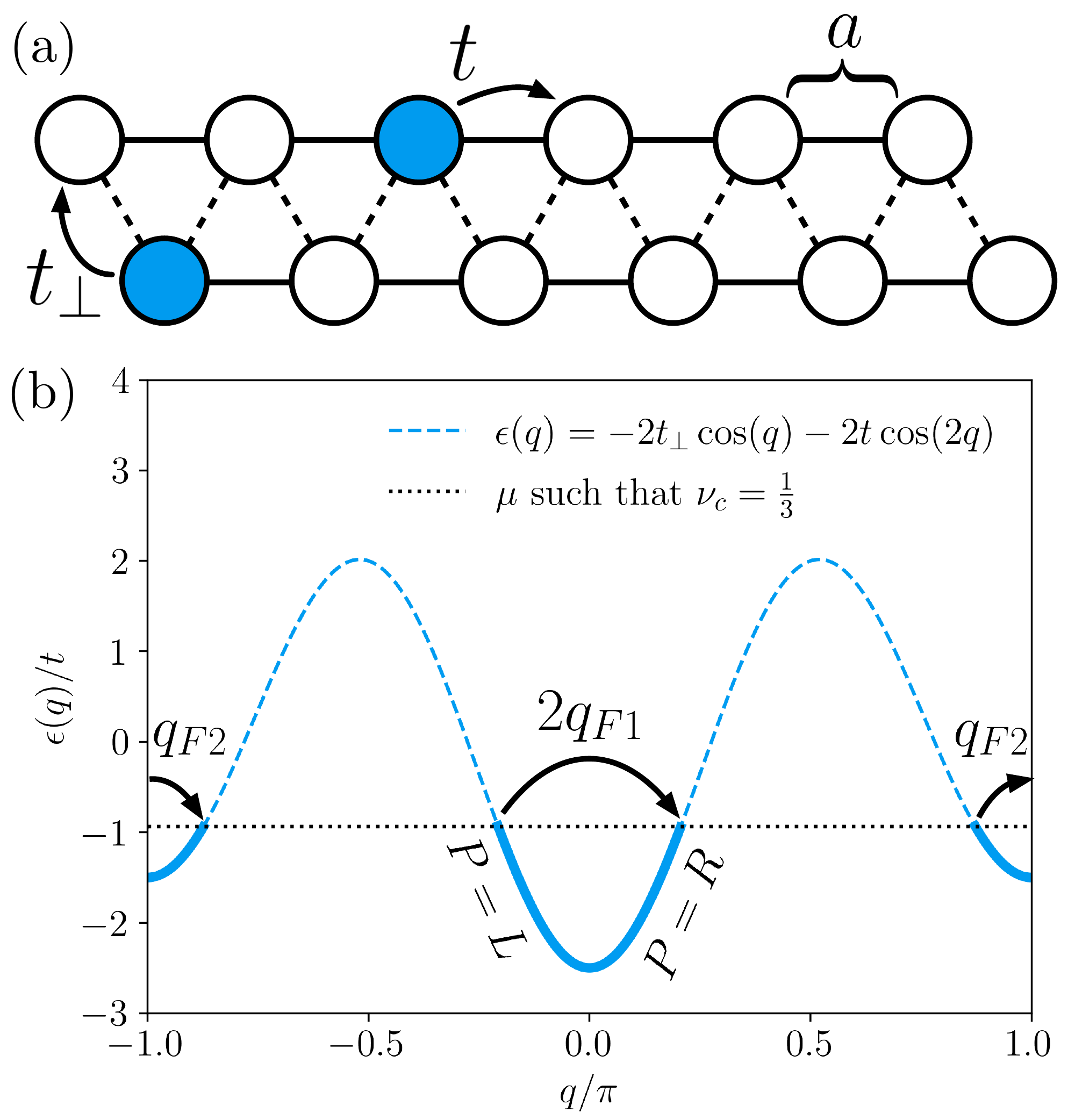}
    \caption{Introduction to the triangle strip model: (a) The real space picture of the model, with charges indicated by sites filled with blue. Intra-chain hopping is given by $t$ and inter-chain hopping is given by $t_\perp$. The distance between sites is $a$. Long-ranged repulsion is not shown but is generically present. (b) The dispersion of the model without interactions, where we treat the model as a single chain with NN hopping $t_\perp$ and NNN hopping $t$. Here $q=ka/2$, where $k$ is the Fermi vector along a single chain. The Fermi wave vectors are labeled and the left- and right-moving fermions of the first Fermi surface are labeled, respectively, by $P=L$ and $P=R$.}
    \label{fig:setup}
\end{figure}

A candidate microscopic model for the two-leg triangular strip shown in Fig.~\ref{fig:setup}(a) can be written as the extended Hubbard model considered by \cite{mishmash_continuous_2015}
\begin{align}
H =& -\sum_{r,\alpha}\left[\left(t_\perp c^\dagger_{r,\alpha}c_{r+1,\alpha} + tc^\dagger_{r,\alpha}c_{r+2,\alpha} + \mathrm{h.c.}\right) - \mu n_r\right] \nonumber\\
&+\frac{1}{2}\sum_{r,r'}V(r-r')n_rn_{r'} \label{eqn:extended_Hubbard},
\end{align}
where $c_{r,\alpha}$ is the electron annihilation operator at site $r$ with spin $\alpha$, $n_r \defeq \sum_\alpha c^\dagger_{r,\alpha}c_{r,\alpha}$ is the electron number operator at site $r$, and $\mu$ is tuned so the system is at an electron filling of $\nu_c$. The inter-electron repulsion $V(r-r')$ is taken to extend over a few lattice constants but is otherwise short ranged and will have an overall scale $U$. Note that we are treating the triangular strip as a single chain with nearest-neighbor (NN) hopping $t_\perp$ and next-nearest-neighbor (NNN) hopping $t$. If we neglect the long-ranged repulsion $V$, then the model will have the dispersion shown in Fig.~\ref{fig:setup}(b). If $t_\perp/t = 0$, the chains will decouple and there will be two FSs, each associated with one of the chains. If $t_\perp/t$ is nonzero, but less than some upper bound dependent on the electron filling $\nu_c$, then there will still be two FSs. We will always consider this limit \footnote{If the model had a single Fermi surface it could be bosonized and treated as in \cite{schulz_metal-insulator_1994}.}. We denote the FS wave vectors by $q_{Fa}$ with $a=1,2$, as shown in Fig.~\ref{fig:setup}(b). They will obey the Luttinger sum rule for this system,
\begin{equation}
2q_{F1}+2q_{F2} = 2\pi \nu_c \pmod{2\pi} \label{eqn:Luttinger},
\end{equation}
(where $\nu_c$ is the filling {\em per spin}) since we have considered a single spin degenerate band. We have set the lattice spacing along the single chain to be $1$. In particular if we consider a filling $\nu_c=1/m$, as we did in two dimensions \cite{musser_theory_2022}, then the Luttinger sum rule means that $2m(q_{F1}+q_{F2}) = 0 \pmod{2\pi}$. This rule will continue to hold even when interactions are turned on.

We introduce this microscopic model merely as a guide to intuition. We ultimately want to understand the universal properties of the metal to WM insulator transition  which should not depend on the microscopic details (except for a finite number of non-universal parameters that enter the low-energy effective theory). In order to do this we will isolate the low-energy degrees of freedom by bosonizing the system and working solely with the bosonized action. We partially \footnote{In \cite{mishmash_continuous_2015} the extra factor of $q_{Fa}r$ in Eq.~(\ref{eqn:bosonized_c}) was absorbed into $\theta_{a\alpha}$.} adopt the convention of Ref. \cite{mishmash_continuous_2015} and expand the electron operator in terms of slowly varying continuum fields at the four Fermi points shown in Fig.~\ref{fig:setup}(b):
\begin{equation}
c_{Pa\alpha} = \eta_{a\alpha} e^{iPq_{Fa}r} e^{i(\phi_{a\alpha} + P\theta_{a\alpha})} \label{eqn:bosonized_c},
\end{equation}
where fluctuations in $\phi_{a\alpha}$ ($\theta_{a\alpha})$ correspond to phase (charge) fluctuations of the $a$th FS with spin $\alpha$. Note that here $P = L (R)$ refers to left (right) moving fermions of a given FS. These fields are canonically conjugate as described in Eqs.~(\ref{eqn:comm_relations}) and the $\eta_{a\alpha}$s are anti-commuting Klein factors which are added to ensure that $c_{Pa\alpha}$ have the correct anticommutation relations.

We further adopt the convention of Ref. \cite{mishmash_continuous_2015} by defining charge and spin modes for each band:
\begin{equation}
\theta_{a\rho/\sigma} \defeq \frac{1}{\sqrt{2}}\left(\theta_{a\uparrow}\pm \theta_{a\downarrow}\right) \label{eqn:defn_charge_spin},
\end{equation}
along with total and relative combinations of the two bands:
\begin{equation}
\theta_{\mu \pm} \defeq \frac{1}{\sqrt{2}}\left(\theta_{1\mu} \pm \theta_{2\mu}\right) \label{eqn:defn_total_relative},
\end{equation}
where $\mu = \rho,\sigma$. The various combinations of the $\phi_{a\alpha}$ are defined analogously. These transformations will then preserve the commutation relations of Eqs.~(\ref{eqn:comm_relations}). With this convention, the total density of electrons can be seen to be
\begin{equation}
\rho(r) = \sum_{P,a,\alpha} c^\dagger_{Pa\alpha} c_{Pa\alpha} = 2\nu_c + \frac{2}{\pi}\partial_r \theta_{\rho+}(r),
\end{equation}
where we used Eq.~(\ref{eqn:Luttinger}) to insert $2\nu_c$. The total charge density of the system is $2\nu_c$, so we see that $\theta_{\rho+}$ represents the long-wavelength fluctuations of the total charge density. Thus any insulating phase of the triangular strip will be characterized by a $\theta_{\rho+}$ which is gapped out.

\section{Metal to WM insulator transition at filling $\nu_c=1/m$}
\label{sec:metal_to_WM}

We have seen that upon bosonization there are generically two charge modes and two spin modes. The metallic state of the bosonized model will leave all of these gapless. In the literature, this is denoted by C$\alpha$S$\beta$, with $\alpha=2$ the number of gapless charge modes and $\beta=2$ the number of gapless spin modes \cite{balents_weak-coupling_1996}. If we now want to promote a transition into an insulating state, then it is necessary to gap out the total charge fluctuation.

In analogy with the case of a one-dimensional chain \cite{schulz_metal-insulator_1994}, we expect that the operator which gaps out the total charge fluctuations will be related to the umklapp operator at filling $\nu_c=1/m$. This operator will transfer a total momentum of $m(2q_{F1} + 2q_{F2}) = 0 \pmod{2\pi}$ and can thus be written schematically as
\begin{align}
\sc{O}_{u} =& \left(\sum_{\alpha} c^\dagger_{R1\alpha}c_{L1\alpha}\right)^m\left(\sum_{\alpha} c^\dagger_{R2\alpha}c_{L2\alpha}\right)^m \nonumber \\
=& \left(\sum_\alpha e^{-2i(q_{F1}r + \theta_{1\alpha})}\right)^m\left(\sum_\alpha e^{-2i(q_{F2}r + \theta_{2\alpha})}\right)^m \nonumber\\
%=& 4^m e^{-\sqrt{2}im (\theta_{1\rho}+\theta_{2\rho})} \cos^m\left(\sqrt{2}\theta_{1\sigma}\right) \cos^m\left(\sqrt{2}\theta_{2\sigma}\right) \nonumber \\
\sim& e^{-2im\theta_{\rho+}}\cos^m\left(\sqrt{2}\theta_{1\sigma}\right) \cos^m\left(\sqrt{2}\theta_{2\sigma}\right) \nonumber \\
\sim& e^{-2im\theta_{\rho+}}\left[\cos\left(2\theta_{\sigma+}\right) + \cos\left(2\theta_{\sigma-}\right)\right]^m\label{eqn:umklapp_operator}.
\end{align}
If we now define 
\begin{align}
\sc{S}_u \defeq& g_u \int \td x \td \tau \left(\sc{O}_u + \sc{O}^\dagger_u\right) \label{eqn:umklapp_pert}\\
\sim& g_u \int \td x \td \tau \ \cos\left(2m\theta_{\rho+}\right)\left[\cos\left(2\theta_{\sigma+}\right) + \cos\left(2\theta_{\sigma-}\right)\right]^m \nonumber,
\end{align}
then this term is consistent with all symmetries of the model, as discussed in Appendix \ref{app:operators_consistent} and must be added to the action. If $\sc{S}_u$ is relevant, then $\theta_{\rho+}$ will be pinned and develop an expectation value and a gap to excitations. The umklapp term is therefore a possible candidate for driving a transition to an insulating state; indeed, we will later explicitly show that generically this term must be what drives an insulating transition.

Gapping out the fluctuations of $\theta_{\rho+}$ will also lead to the development of charge-density wave (CDW) order. If $\theta_{\rho+}$ is pinned by umklapp scattering, we should generically expect that any operator which depends solely on $\theta_{\rho+}$ will develop an expectation value. In particular, consider the operator
\begin{equation}
\sc{O}_{\mathrm{CDW}} = c^\dagger_{R1\uparrow}c^\dagger_{R1\downarrow}c^\dagger_{R2\uparrow}c^\dagger_{R2\downarrow}c_{L1\uparrow}c_{L1\downarrow}c_{L2\uparrow}c_{L2\downarrow} \label{eqn:defn_OCDW},
\end{equation}
which has a bosonized form given by
\begin{equation}
\sc{O}_{\mathrm{CDW}}\sim e^{4i\theta_{\rho+}}e^{2\pi i (2\nu_c)r}.
\end{equation}
We show in Appendix \ref{app:CDW_enumerated} that all operators which depend only on $\theta_{\rho+}$ will simply be powers of $\sc{O}_{\mathrm{CDW}}$. This operator can be seen to be precisely the order parameter for CDW order with real-space periodicity of $1/(2\nu_c) = m/2$. Thus by gapping out $\theta_{\rho+}$ we have not only gapped out the fluctuations of the total charge, thus making the system an insulator, but have given finite expectation value to a CDW order parameter with periodicity equivalent to the total charge. This is precisely our definition of a transition from a metal to a WM insulator, as discussed in the introduction and illustrated in Fig.~\ref{fig:WM_defn}.

The relevance or irrelevance of $\sc{S}_u$ is controlled by the sign of
\begin{equation}
\Delta[g_u] = 2 - \Delta[\sc{O}_u].
\end{equation}
Assuming that we reside on the metallic side of the transition, the fixed point theory of the system will generically be quadratic and have the scaling dimensions,
\begin{equation}
\Delta\left[e^{i\beta \theta_{\rho+}}\right] = \frac{\beta^2 K_{\theta_{\rho+}}}{4} \text{ and } \Delta\left[e^{i\gamma \theta_{\sigma\pm}}\right] = \frac{\gamma^2 K_{\theta_{\sigma\pm}}}{4} \label{eqn:scaling_necess},
\end{equation}
where $K_{\theta_{\mu\pm}}$ ($K_{\phi_{\mu\pm}}$) is defined to be the scaling dimension of $e^{2i\theta_{\mu \pm}}$ ($e^{2i\phi_{\mu\pm}}$) for $\mu = \rho,\sigma$. The $\mathrm{SU}(2)$ invariance of the theory will mean that the Hamiltonian is diagonal in the spin sector and $K_{\theta_{\sigma\pm}} = 1 = K_{\phi_{\sigma\pm}}$. Note that the generic metallic Hamiltonian will not be diagonal in the charge sector, and so we will generically have $K_{\theta_{\rho \pm}} \neq K_{\phi_{\rho\pm}}^{-1}$. If the charge sector Hamiltonian \textit{is} diagonal, then we will follow the convention of Ref. \cite{mishmash_continuous_2015} and set $K_{\theta_{\rho\pm}}\equiv K_{\rho\pm}$ and $K_{\phi_{\rho \pm}} \equiv K_{\rho\pm}^{-1}$. This is discussed in much greater detail in Appendix \ref{app:theory_and_scaling}. Thus, to identify $\Delta[\sc{O}_u]$ we must expand the spinful terms to their lowest order. For even $m$, the most relevant contribution of these terms will simply be a constant, while if $m$ is odd the most relevant contribution will be $\cos(2\theta_{\sigma\pm})$. Then we see from Eqs.~(\ref{eqn:scaling_necess}) that
\begin{equation}
\Delta[g_u] = \begin{cases} 2 - m^2K_{\theta_{\rho+}} &\mbox{if } m \text{ even}\\ 1 - m^2K_{\theta_{\rho+}} &\mbox{if } m \text{ odd,}\end{cases}
\end{equation}
which comes from the most relevant parts of $\sc{S}_u$:
\begin{equation}
\sc{S}_u \sim \int \td x \td \tau \ \cos(2m\theta_{\rho+}) \times \begin{cases} 1 &m \in 2\bb{Z},\\ \cos(2\theta_{\sigma\pm})  &\mbox{else.}\end{cases}
\end{equation}

If we then start from the metallic (C2S2) state and drive an insulating transition via $\sc{S}_u$ by tuning $K_{\theta_{\rho+}}$ until $\Delta[g_u] < 0$, the equations above suggest we must treat fillings $\nu_c=1/m$ very differently depending on the parity of $m$. For even $m$, the most relevant term in $\sc{S}_u$ will gap out only the total charge, while for odd $m$ the spin excitations will both be gapped out. This means that $\sc{S}_u$ describes a transition from the metal C2S2 to C1S2 when $m$ is even, but a transition from C2S2 to C1S0 when $m$ is odd \cite{balents_weak-coupling_1996}. 

The properties of these phases have been analyzed by Sheng \textit{et al.} for the case of $\nu_c=1/m$ with $m=2$, i.e., without translational symmetry breaking \cite{sheng_spin_2009}. There it was found that the C1S2 phase was stable and provided a one-dimensional example of the spinon FS state. Additionally, if interactions in the spin sector became marginally relevant, then this C1S2 phase could flow to a C1S0 state where $\theta_{\rho+}$ and $\theta_{\sigma\pm}$ were pinned. This state exhibited period-$2$ VBS order, but showed power-law correlations in the VBS order parameter at the incommensurate wave vectors $2q_{F1}$ and $2q_{F2}$. The C1S0 state we describe will have similar properties,  exhibiting period-$m$ VBS order, but showing power-law correlations in the VBS order parameter at $2q_{F1}$ and $2q_{F2}$.

Of course, there could be some $m$-independent term in the action which is more relevant than $\sc{S}_u$ and the distinction between even and odd $m$ will hence be avoided. We will show in the next section that this is not the case, but for now we make an argument that these fillings must indeed be treated quite differently by considering $m=6$ and $m=3$.

\begin{figure}
    \centering
    \includegraphics[width=\columnwidth]{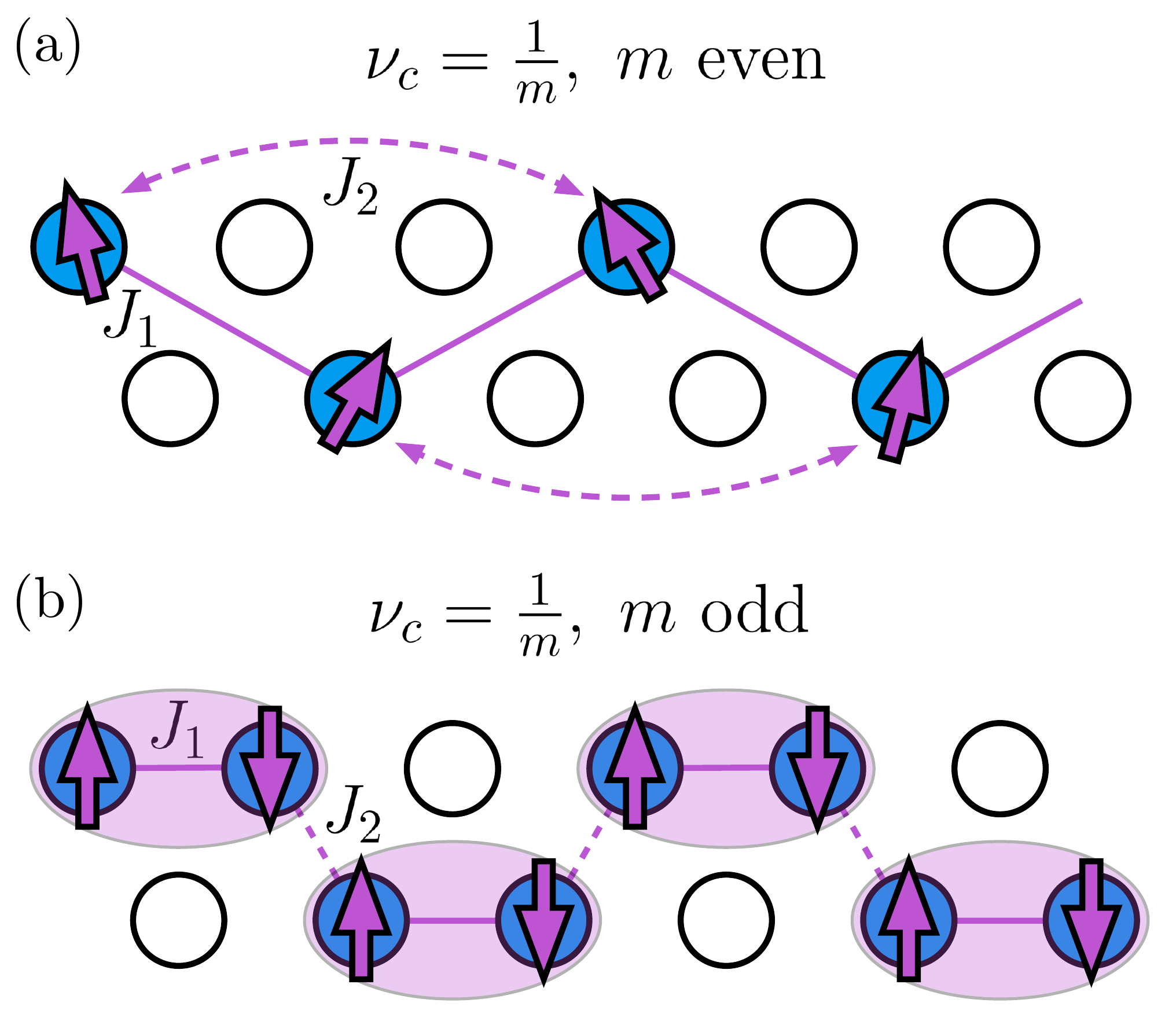}
    \caption{Behavior of the Wigner-Mott insulator deep in the insulating phase: (a) The system with $\nu_c=1/6$. We expect the charge order shown, with the filled circles indicating electron charge. The ground state of the system is then a spin$-1/2$ chain with NN spin exchange $J_1$, shown as a magenta line, and NNN spin exchange $J_2$, shown as a dotted magenta line. Deep in the insulating phase $J_1/J_2\gg 1$, leading to gapless spins. (b) For $\nu_c = 1/3$, we expect the charge order shown here. The ground state is then a single spin$-1/2$ chain with alternating spin exchange scales $J_1,J_2$ denoted by a solid and dotted magenta line, respectively. Here we consider $J_1/J_2>1$ so singlet formation is encouraged along the solid magenta lines; these singlets are indicated with magenta ellipses.}
    \label{fig:even_odd}
\end{figure}

Consider the case of $\nu_c = 1/6$ deep in the insulating regime, illustrated in Fig.~\ref{fig:even_odd}(a). We expect that the charge-ordering pattern will be as shown for the microscopic model given in Eq.~(\ref{eqn:extended_Hubbard}) based on the observed and predicted $\sqrt{3}\times \sqrt{3}$ charge ordering on the full triangular lattice in two dimensions \cite{xu_correlated_2020, li_imaging_2021}. With this charge ordering, the system can be treated as a single spin-$1/2$ chain with NN spin exchange $J_1$ and NNN spin exchange $J_2$. Their ratio will be roughly $J_1/J_2 \sim t_\perp^2 U^2/t^4$, where $U$ is an overall scaling factor for the long-ranged repulsion $V(r-r')$ in Eq.~(\ref{eqn:extended_Hubbard}). Regardless of the ratio of $t_\perp/t$, as $U/t$ is tuned deep into the insulating regime $J_1/J_2\rightarrow \infty$. In this limit, the spins remain gapless \cite{chitra_density-matrix_1995, white_dimerization_1996}. Thus, we might expect that any transition from the metal into the insulating state with charge order shown in Fig.~\ref{fig:even_odd}(a) will not gap out the spins.

Now consider the case of $\nu_c=1/3$ deep in the insulating regime. We can make a particle-hole transformation of the above case, which will turn the $\nu_c=1/6$ filled ladder into the $\nu_c=1/3$ filled ladder. Since the Coulomb interaction is particle-hole symmetric, we should thus expect the charge ordering shown in Fig.~\ref{fig:even_odd}(b). Again, we may treat the system deep in the insulating regime with this charge ordering pattern as a single spin$-1/2$ chain. However, this chain will, to leading order, only have alternating NN hopping. The NN hoppings will scale as $J_1 \sim t^2/U$ and $J_2\sim t_\perp^2/U$. These couplings will be different unless the system is fine tuned, and the spins will thus be gapped out \cite{hida_crossover_1992}. Thus, we might also expect that any transition from the metal into the insulating state with this charge order will also gap out the spins. Note also that this state will indeed exhibit the period-$m$ VBS order expected for the C1S0 state where $\theta_{\rho+}, \theta_{\sigma\pm}$ are pinned.

These examples provide heuristic reasons to suspect that the spins must also be gapped out at the metal to WM insulator transition when $\nu_c = 1/m$ with $m$ odd, but ultimately we must consider the full bosonized theory to address this question.

\section{Transitions out of the metallic theory}
\label{sec:dominant_umklapp}

We now want to investigate more thoroughly how to tune from the metallic C2S2 theory to a different fixed point theory where $\theta_{\rho+}$ is gapped. As we do so, we want to be careful to avoid any intervening phase where some combination of the four modes are gapped out, but $\theta_{\rho+}$ is not. Such an intervening phase would preclude the kind of direct metal to WM transition shown in Fig.~\ref{fig:WM_defn}. We thus need to consider all possible terms that might be added to the generic Luttinger liquid action for the C2S2 metal and show that these terms will not become relevant before a term that gaps out the total charge. In doing so, we will see that for many, though not all, transitions the umklapp operator, we have described is always the most relevant as the critical point is approached from the metallic side. For these transitions, the physics described in the previous section will continue to hold.

We begin by enumerating the possible operators to add to the metallic theory; these will be all operators consistent with the symmetries of the C2S2 metal. These symmetries can be seen to be particle number conservation, conservation of total spin, $\mathrm{SU}(2)$ invariance, time reversal invariance, and the space-group symmetries of the triangular lattice strip in Fig.~\ref{fig:setup}(a). The space-group symmetries will imply conservation of crystal momentum in addition to the point-group symmetries of the strip, which consist only of reflection about lattice sites. In Appendix \ref{app:operators_consistent}, we work out the most general operator consistent with all of these symmetries. 

An important symmetry for our purposes will be the conservation of crystal momentum. The total charge mode $\theta_{\rho+}$ has an accompanying momentum of $q_{F1} + q_{F2}$, while the relative charge mode has an accompanying momentum of $q_{F1} - q_{F2}$. While $q_{F1} + q_{F2}$ is constrained to be a rational multiple of $2\pi$ by the Luttinger sum rule, Eq.~(\ref{eqn:Luttinger}), $q_{F1} - q_{F2}$ is incommensurate in the absence of fine tuning. Thus, the conservation of crystal momentum will mean that any operator allowed to be added to the C2S2 theory cannot contain terms with the relative charge mode $\theta_{\rho-}$. With these constraints, we show in Appendix \ref{app:most_relevant_operators} that the most relevant operators in the metallic C2S2 theory will be the umklapp operator and the operators $\cos(2\phi_{\rho-})\cos(2\phi_{\sigma-})$ and $\cos(2\phi_{\rho-})\cos(2\theta_{\sigma\pm})$, where these must be added together in an $\mathrm{SU}(2)$ invariant way. The latter operator is the four-fermion $W$ operator discussed in Ref. \cite{sheng_spin_2009}, which mixes right- and left-moving fermions at different Fermi points. We will henceforth refer to it as $\sc{O}_{\mathrm{W}}$.

Suppose now that our theory is diagonal in the charge sector. Then we can tune $K_{\rho+} = K_{\theta_{\rho+}}$ to zero to drive an insulating transition via the umklapp operator. Since we do not alter $K_{\phi_{\rho-}} = K_{\rho-}^{-1}$, then the $\sc{O}_{\mathrm{W}}$ term will stay irrelevant, as it must have been irrelevant by definition in the C2S2 metallic theory. As these are the two most relevant terms, we conclude that this describes a direct metal to WM transition in the two-leg ladder. Of course, a given microscopic theory is not likely to have a charge sector that is diagonal. Nonetheless, as long as $K_{\phi_{\rho-}}$ does not decrease as $K_{\theta_{\rho+}}$ is tuned to zero, then a direct metal to WM transition will be obtained. If, on the other hand, $K_{\phi_{\rho-}}$ does decrease as $K_{\theta_{\rho+}}$ is tuned to zero, then the transition that occurs will be determined by whether the umklapp or the $\sc{O}_{\mathrm{W}}$ term becomes relevant first. In the first case, a direct metal to Wigner-Mott transition will still occur. In the second case, the transition would be into the C$1$S$0$ state with the total charge remaining gapless. As $K_{\theta_{\rho+}}$ continues to decrease, there will then be a second transition to the C$0$S$0$ insulating state. Given that we have now demonstrated (in the diagonal theory) that a direct metal to WM transition is possible, we now consider subsequent transitions to the fully gapped C0S0 state.

After the umklapp operator becomes relevant, the even $m$ system will flow to the C1S2 state pictured in Fig.~\ref{fig:WM_defn}(a). One can redo the arguments in Appendix \ref{app:most_relevant_operators} to show that the most relevant possible $\mathrm{SU}(2)$ invariant operator containing $\phi_{\rho-}$ will again be the $\sc{O}_{\mathrm{W}}$ term. We should thus expect that any subsequent transition where $\phi_{\rho-}$ is pinned will also gap out both spin modes if $m$ is even. Alternatively, the spin terms may become marginally relevant and gap out the spin modes before $\phi_{\rho-}$ is pinned. The possible transitions out of the C1S2 intermediate state are then either to the fully gapped C0S0 state or to the C1S0 state where the relative charge mode remains gapless. This is indeed what was found by Sheng \textit{et al}.~\cite{sheng_spin_2009} for the case of $m=2$. We note that this means the strong coupling picture shown in Fig.~\ref{fig:even_odd}(a) will not generically be an accessible phase via the transitions we are discussing, since it has both charge modes gapped but a gapless spin mode. To access this phase, it is necessary for $q_{F_1}-q_{F_2}$ to be renormalized to a commensurate wavevector so $\theta_{\rho-}$ can be pinned by another umklapp operator.

Finally, after the umklapp operator becomes relevant for $m$ odd, the system will flow to the C1S0 state with a gapless relative charge mode. This is shown as the intermediate phase in Fig.~\ref{fig:WM_defn}(b). Alternatively, as discussed above, this state can be accessed for $m$ even by the spin terms becoming marginally relevant before $\phi_{\rho-}$ is pinned. The transition out of this state to the fully gapped C0S0 state will be driven by terms proportional to $\cos(2\phi_{\rho-})$, where the coefficient of proportionality will be a function of the pinned total charge and spin modes.

Ultimately, the transition we are interested in is the metallic C2S2 to WM transition; we will thus not concern ourselves further with the transition to the fully gapped state.

\section{Physical properties of the transition}

We discuss several properties of the metal to WM transition for the case of $m$ even and $m$ odd. In each case, we will discuss the RG flow of the theory and the scaling of any gaps in the system, operators which acquire an expectation value at the critical point, and Green's functions. We will also discuss the behavior of the compressibility and spin susceptibility.

\subsection{Even $m$}

The transition in this case is from the C2S2 metal to the C1S2 state and is driven by the umklapp operator $\cos(2m\theta_{\rho+})$. We therefore want to study the properties of the theory given by
\begin{equation}
\sc{L} = \sc{L}^\rho_{\mathrm{C2S2}} + \sc{L}^\sigma_{\mathrm{C2S2}} + 2g_u\cos(2m\theta_{\rho+}) \label{eqn:even_m_full_theory},
\end{equation}
with the preceding terms given by Eqs.~(\ref{eqn:C2S2_charge}) and (\ref{eqn:C2S2_spin}). The critical properties of this transition were studied by Mishmash \textit{et al}.~\cite{mishmash_continuous_2015} for $m=2$. If $\theta_{\rho+}$ in our theory is rescaled to $2\theta_{\rho+}/m$, the critical properties will then be identical. The transition is Kosterlitz-Thouless(KT)-like, with an additional complication arising because the $\sc{L}^\rho_{\mathrm{C2S2}}$ theory is not necessarily diagonal in the $\rho\pm$ basis. Nonetheless, near the critical point a change of variables reveals that the transition is a simple KT transition, with the relevance of $g_u$ being controlled by $2-m^2K_{\theta_{\rho+}}$ and the non-diagonal RG flow being entirely controlled by the flow of $K_{\theta_{\rho+}}$ and $g_u$. Thus, for $K_{\theta_{\rho+}} > m^2/2$, the theory flows to a C2S2 fixed point, while for $K_{\theta_{\rho+}} < m^2/2$ $g_u$ flows to infinity, and $K_{\theta_{\rho+}}$ flows to zero. The velocities of the diagonal modes of the $\sc{L}^\rho_{\mathrm{C2S2}}$ theory are not renormalized by the flows, nor is the spin sector.

The compressibility, which measures the response of the theory to a $-\mu \int \td r \td \tau \ \rho = - \mu \int\td r \td \tau \ (2\partial_r \theta_{\rho+}/\pi)$ term in the action, will thus drop discontinuously to zero on the insulating side where $\theta_{\rho+}$ is pinned. At the critical point, the value of $\kappa$ will be nonuniversal. Indeed, if the theory were to be diagonal in the $\rho\pm$ basis, $\kappa = 4K_{\theta_{\rho+}}/\pi v_{\rho+}$, where $v_{\rho+}$ will take a non-universal value at the critical point. Similarly, the spin susceptibility is given by $\chi = 4K_{\theta_{\sigma+}}/\pi v_{\sigma+} = 4/\pi v_{\sigma+}$ and therefore evolves smoothly across the transition.

Near the critical point, the correlation length on the insulating side will follow the usual KT form
\begin{equation}
\xi^{-1} \sim \exp\left(-\frac{C}{\sqrt{2-m^2K_{\theta_{\rho+}}}}\right),
\end{equation}
where $C$ is some non-universal constant. We can expect the charge gap to scale like $\Delta_c \sim \hbar v_{\rho+}\xi^{-1}$; since $v_{\rho+}$ evolves smoothly across the transition $\Delta_c \sim \xi^{-1}$ near the critical point.

The behavior of the Green's function, $\langle c^\dagger_\alpha(x) c_\alpha(0)\rangle$ across this transition was discussed in Ref.~\cite{mishmash_continuous_2015}. There they note that the electron operator will contain the conjugate field to $\theta_{\rho+}$, $\phi_{\rho+}$. Once $\theta_{\rho+}$ is pinned, this will fluctuate wildly and cause it to decay exponentially at all wave vectors. This same behavior will also be present in our case. Thus, on the metallic side, we expect the Green's function to decay as a power law, while on the insulating side it will decay exponentially. At the critical point, where $K_{\theta_{\rho+}} = 2/m^2\neq 0$, the Green's function will still exhibit power-law decay.

The transition for $m\neq 2$ has the additional feature that the operator $\sc{O}_{\mathrm{CDW}} \sim e^{4i\theta_{\rho+}} e^{4\pi i r/m}$ will break translational symmetry and develop an expectation value on the insulating side. We make a standard argument to relate its scaling to the scaling dimension of $\sc{O}_{\mathrm{CDW}}$. Consider the correlator $\langle \sc{O}_{\mathrm{CDW}}(x) \sc{O}_{\mathrm{CDW}}(0) \rangle$. Just on the insulating side where $K_{\theta_{\rho+}}$ has not renormalized to zero, we expect that this correlator will continue to scale as $|x|^{-2\Delta_{\mathrm{CDW}}}$ along with some possible logarithmic corrections arising from the marginally irrelevant umklapp operator \cite{giamarchi_quantum_2003}. However, fluctuations of $\theta_{\rho+}$ should be suppressed for length scales larger than $\xi$. We may thus conclude that $\langle \sc{O}_{\mathrm{CDW}}\rangle \sim \sqrt{\langle \sc{O}_{\mathrm{CDW}}(\xi)\sc{O}_{\mathrm{CDW}}\rangle}$. Then up to some logarithmic corrections $\langle \sc{O}_{\mathrm{CDW}}\rangle \sim \xi^{-4K_{\theta_{\rho+}}}$ since $\sc{O}_{\mathrm{CDW}} \sim e^{4i\theta_{\rho+}}e^{2\pi i(2\nu_c r)}$. For the sake of completeness, we compute the logarithmic corrections to $\langle \sc{O}_{\mathrm{CDW}}(\xi)\sc{O}_{\mathrm{CDW}}\rangle$ in the theory diagonal in the $\rho\pm$ basis in Appendix \ref{app:evenm_correlations} and find that
\begin{equation}
\langle \sc{O}_{\mathrm{CDW}}\rangle \sim \exp\left(-\frac{8C/m^2}{\sqrt{2 - m^2K_{\rho+}}}\right)\left(2-m^2K_{\rho+}\right)^{-4/m^2}.
\end{equation}
The non-diagonal theory should display identical scaling once the appropriate change of variables is made near the critical point.

\subsection{Odd $m$}

In the case of odd $m$, the umklapp operator now contains spin terms. We thus want to study the properties of the theory given by
\begin{align}
\sc{L} =& \sc{L}^\rho_{\mathrm{C2S2}} + \sc{L}^\sigma_{\mathrm{C2S2}} + 4g_1\cos(2m\theta_{\rho+})\cos(2\theta_{\sigma+})\\
&+ 4g_2\cos(2m\theta_{\rho+})\cos(2\theta_{\sigma-})\\
&+ 4g_3\cos(2m\theta_{\rho+})\cos(2\phi_{\sigma-}).
\end{align}
Note that this is not the same as Eq.~(\ref{eqn:umklapp_pert}), which had $g_1=g_2=g_u$ and $g_3=0$. This is because we want to consider the possibility that the coefficients of the two umklapp terms will renormalize differently and to include the most relevant terms containing the total charge mode consistent with the symmetries of $\sc{L}_{\mathrm{C2S2}}$, as discussed in Appendix \ref{app:most_relevant_operators}. Here $\mathrm{SU}(2)$ symmetry is not manifest, so we must ensure that the $g_1,g_2,g_3$ coefficients are chosen so it is maintained by the theory. To do this, we note that, as in Ref.~\cite{mishmash_continuous_2015}, we may write the bosonized spin operators at wave vector $2q_{F_a}$, $\vec{S}_{2q_{F_a}} = c^\dagger_{Ra\alpha}\vec{\sigma}_{\alpha\beta}c_{La\beta}/2$, as
\begin{align}
S^x_{2q_{F_a}} =& -i\eta_{a\uparrow}\eta_{a\downarrow} e^{i\theta_{\rho+}}e^{i\sigma_a\theta_{\rho-}}\sin(\sqrt{2}\phi_{a\sigma})\\
S^y_{2q_{F_a}} =& -i\eta_{a\uparrow}\eta_{a\downarrow}e^{i\theta_{\rho+}}e^{i\sigma_a \theta_{\rho-}}\cos(\sqrt{2}\phi_{a\sigma})\\
S^z_{2q_{F_a}} =& -e^{i\theta_{\rho+}}e^{i\sigma_a \theta_{\rho-}}\sin(\sqrt{2}\theta_{a\sigma}).
\end{align}
Here $\theta_{\sigma\pm} = (\theta_{1\sigma}\pm \theta_{2\sigma})/\sqrt{2}$. We now consider a $270^{\circ}$ rotation about the $x$ axis which takes $S^z\rightarrow S^y$ and $S^y\rightarrow -S^z$ while leaving $S^x$ invariant. This will take $\sin(\sqrt{2}\theta_{a\sigma})\rightarrow i\eta_{a\uparrow}\eta_{a\downarrow}\cos(\sqrt{2}\phi_{a\sigma})$ and $i\eta_{a\uparrow}\eta_{a\downarrow}\cos(\sqrt{2}\phi_{a\sigma})\rightarrow -\sin(\sqrt{2}\theta_{a\sigma})$ while leaving $\sin(\sqrt{2}\phi_{a\sigma})$ invariant. Further, we know that $\cos(\sqrt{2}\theta_{a\sigma})$ must be invariant under this transformation because $\delta n_{2q_{F_a}} \propto \cos(\sqrt{2}\theta_{a\sigma})$ and the density operator is invariant under all spin rotations. The quadratic part of the theory is trivially invariant under this rotation. Expanding the $\sigma\pm$ basis in terms of the $a\sigma$ basis allows us to see that $(g_1-g_2)^2 = g_3^2$ for the interacting part of the action to be invariant. The squared exponent must be included to deal with the sign arising from the choice of Majorana sector. If the C2S2 theory is initially perturbed by the umklapp operator, then $g_1(l=0) = g_u = g_2(l=0)$ and thus $g_3(l=0)=0$ initially. But then $g_3(l)=0$ for the whole flow, since $\td g_3/\td l \propto g_3$. Thus, $\mathrm{SU}(2)$ invariance requires that the C2S2 phase, when initially perturbed by the umklapp term, will have a theory as above but with $g_3=0$. If there were some other perturbation such that $g_1-g_2\neq 0$, then this additional term would need to be considered.

Even without a $g_3$ term, we might expect that this transition will no longer be of a simple KT type, since there are now two equally relevant terms driving the transition. In particular, it might be the case that the spin gap, which is now opened at the same critical point, scales differently than the charge gap. To simplify our treatment of the transition, we will ignore a number of subtleties. We will treat the $\sc{L}_{\mathrm{C2S2}}$ theory as being diagonal in the $\rho\pm$ basis. The RG flow near the critical point should be unaffected by this choice, as we can always linearize near it and consider only deformations along the $K_{\theta_{\rho+}}$ direction. This is analogous to the approach taken by Mishmash \textit{et al}.~\cite{mishmash_continuous_2015} in the $m=2$ case. We will also work with the theory where all of the spin and charge velocities are identical. This makes it easier to compute the RG flows in real space, but again should capture the RG flows we care about since the velocities should be expected to evolve smoothly across the critical point. With these simplifications, we treat the RG flows in Appendix \ref{app:oddm_correlations}. 

We discuss the flows in more detail in the Appendices, but we note that because $K_{\sigma+} = 1 = K_{\sigma-}$ and $g_1=g_u=g_2$ initially, they will maintain this equality throughout the flow. The problem then reduces once more to a simple KT-like transition which is driven by a single perturbing term $g_u$, with $K_{\rho+}$ and $K_{\sigma\pm}$ being driven to zero in a similar way. On the separatrix, where the flow is the simplest their flows will take the form
\begin{align}
x_{\rho+}(l) =& \frac{2}{3}\frac{x_{\rho+,0}}{1-x_{\rho+,0}l} + \frac{1}{3}x_{\rho+,0}, \text{ and}\\
x_{\sigma\pm}(l) =& \frac{1}{3}\frac{x_{\rho+,0}}{1-x_{\rho+,0}l} - \frac{1}{3}x_{\rho+,0},
\end{align}\
where $x_{\rho+} = 1-m^2K_{\rho+}$ and $x_{\sigma\pm} = 1-K_{\sigma\pm}$. Here we see that $\td x_{\rho+}/\td l = 2\td x_{\sigma\pm}/\td l$; this is because there are two terms proportional to $g_u$ which attempt to pin $\theta_{\rho+}$, while there is only a single term for each $\theta_{\sigma\pm}$ so its flow is twice as fast. Despite this, we note that for $x_{\rho+,0}$, the $l$ required for $x_{\rho+}$ and $x_{\sigma\pm}$ to become order one will be $1/x_{\rho+,0}$ to leading order for both. We therefore conclude that the charge and spin gaps must scale identically as
\begin{equation}
\Delta_c \sim \Delta_s\sim \exp\left(-\frac{C}{\sqrt{2-m^2K_{\rho+}-K_{\sigma+}}}\right),
\end{equation}
where $C$ is some non-universal number. Note that we have not set $K_{\sigma+} = 1$ here to account for its value being smaller than one on the insulating side with a spin gap.

Just as in the even $m$ case, the Green's function will continue to have power-law decay on the metallic side and at the critical point, while it will decay exponentially on the insulating side. The behavior of the compressibility will also be analogous; it will drop discontinuously to zero on the insulating side and take a non-universal value at the critical point. In this case, the spin susceptibility will behave identically, as the spins are gapped on the insulating side.

We discuss the scaling of the expectation value of the CDW operator near the critical point in Appendix \ref{app:oddm_correlations}. There, we find that
\begin{align}
\langle \sc{O}_{\mathrm{CDW}}\rangle \sim& \exp\left(-\frac{4C/m^2}{\sqrt{2-m^2K_{\rho+}-K_{\sigma+}}}\right)\\
&\times\left(2-m^2K_{\rho+}-K_{\sigma+}\right)^{-8/3m^2},
\end{align}
where the presence of the $8/3m^2$ in the exponent of the logarithmic correction is due to the $2/3$ factor multiplying the diverging piece of $x_{\rho+}$.

Finally, we discuss the scaling of the other important symmetry breaking order on the insulating side; the period-$m$ VBS order. This will lead to an expectation value for $\sc{B}_{2\pi/m}$, where $\sc{B}_Q$ is the VBS order parameter at wave vector $Q$. By analogy, with Ref.~\cite{mishmash_continuous_2015} we expect that the bosonized form of this operator will be given by
\begin{equation}
\sc{B}_{2\pi/m} \sim [\cos(2\theta_{\sigma+})+\cos(2\theta_{\sigma-})]\sin(m\theta_{\rho+}).
\end{equation}
Thus, $\Delta[\sc{B}_{2\pi/m}] = K_{\sigma+} + m^2K_{\rho+}/4$ which is equal to $5/4$ at the critical point. We should thus expect that $\langle \sc{B}_{2\pi/m}\rangle \sim \Delta_c^{5/4}$ (plus logarithmic corrections which we will not compute).

We conclude our discussion by noting that a full treatment of the critical point will include the possibility of marginally relevant terms which couple the spin modes together. These are written as the $\sigma$ terms in Refs.~\cite{sheng_spin_2009, mishmash_continuous_2015}. The most important contribution comes from a term proportional to $\cos(2\theta_{\sigma+})\cos(2\phi_{\sigma-})$. If its coefficient is nonzero, then it will cause $K_{\sigma+}$ and $K_{\sigma-}$ to renormalize to zero at different rates and break the simple KT-like nature of the transition. Nonetheless, none of these additional terms will stop the umklapp term from pinning both spin modes.

\section{Discussion}

In this paper, we analyzed the WM transition in the two-leg triangular ladder. We were able to show using bosonization that a continuous bandwidth-tuned metal to WM transition is possible and is driven by the umklapp operator. An interesting extension of our results would be to study the density-tuned transition, which would allow for us to treat experiments that are tuned through the transition via doping.

Additionally, we were able to describe, in the case of odd denominator fillings, a transition which opened a spin gap at the same critical point. Such a transition has been described in the case of a $d=1$ chain \cite{schulz_metal-insulator_1994}, but its extension to multi-leg ladders may provide a way to tackle a continuous metal-insulator transition in two dimensions which does not exhibit a spinon FS on the insulating side. To this end, it may be interesting to extend our results to higher leg ladders.

Finally, it would be interesting to consider the transition for even denominator fillings at a finite temperature. Deep in the insulating side, the increased distance between charges may mean that the spin exchange scale is substantially renormalized downwards in a non-universal way. Thus, the transition for even denominator fillings at finite temperature should be expected to be a transition between a metal and a spin-incoherent Luttinger liquid \cite{fiete_colloquium_2007} with a remaining gapless degree of freedom encoded in the relative charge. Extending this to two dimensions is less clear, but may be relevant to experiments. Indeed, one may expect that future experiments probing the metal to WM transition of moire-TMDs with filling $\nu_c < 1/2$ will describe a similar metal to spin incoherent liquid transition.

\acknowledgements 
We thank Debanjan Chowdhury for stimulating discussions, and for a previous collaboration on related topics. SM was supported by the National Science Foundation Graduate Research Fellowship under Grant No. 1745302. Any opinions, findings, and conclusions or recommendations expressed in this material are those of the author(s) and do not necessarily reflect the views of the National Science Foundation. T.S. was supported by US Department of Energy Grant No. DE- SC0008739, and partially through a Simons Investigator Award from the Simons Foundation. This work was also partly supported by the Simons Collaboration on Ultra-Quantum Matter, which is a grant from the Simons Foundation (Grant No. 651446, T.S.).

\appendix

\section{Generic $\sc{L}_{\mathrm{C2S2}}$ and scaling dimensions}
\label{app:theory_and_scaling}

We first state the commutation relations. The fields $\theta_{a\alpha}$ and $\phi_{a\alpha}$ are canonically conjugate so
\begin{align}
[\phi_{a\alpha}(r), \phi_{b\beta}(r')] =& 0 = [\theta_{a\alpha}(r), \theta_{b\beta}(r')], \nonumber\\
[\phi_{a\alpha}(r), \theta_{b\beta}(r')] =& i\pi\delta_{ab}\delta_{\alpha\beta}\Theta(r-r') \label{eqn:comm_relations}.
\end{align}

We will now draw heavily from the Appendix of Ref.~\cite{lai_two-band_2010}. The generic action for C2S2 will be quadratic, along with possible marginal corrections, and can be written as
\begin{align}
\sc{L}_{\mathrm{C2S2}} =& \sc{L}^\rho_{\mathrm{C2S2}} + \sc{L}^\sigma_{\mathrm{C2S2}}, \text{ where} \label{eqn:C2S2_decouple}\\
\sc{L}^\rho_{\mathrm{C2S2}} =& \frac{1}{2\pi}\left[\partial_r \vec{\theta}^T_\rho A \partial_r \vec{\theta}_\rho + \partial_r \vec{\phi}^T_\rho B \partial_r \vec{\phi}_\rho\right] + \frac{i}{\pi}\partial_r \vec{\theta}_\rho \cdot \partial_\tau \vec{\phi}_\rho\label{eqn:C2S2_charge},
\end{align}
with $A$ and $B$ some real, symmetric, positive-definite matrices whose entries can be deduced starting from a generic microscopic model \cite{mishmash_continuous_2015}. Here $\vec{\theta}^T_\rho = (\theta_{\rho+}, \theta_{\rho-})$ and $\vec{\phi}^T_\rho = (\phi_{\rho+}, \phi_{\rho-})$ and the time derivative term is enforced by the commutation relations in Eqs.~(\ref{eqn:comm_relations}). The Lagrangian in the spin sector will be somewhat simpler due to $\mathrm{SU}(2)$ invariance:
\begin{align}
\sc{L}^\sigma_{\mathrm{C2S2}} =& \sum_{s=\pm} \frac{v_{\sigma s}}{2\pi}\left[\frac{1}{K_{\sigma s}}\left(\partial_r \theta_{\sigma s}\right)^2 + K_{\sigma s}\left(\partial_r \phi_{\sigma s}\right)^2\right] \nonumber\\
&+ \sum_{s=\pm}\frac{i}{\pi}\left(\partial_r \theta_{\sigma s}\right)\left(\partial_\tau \phi_{\sigma s}\right)\label{eqn:C2S2_spin}.
\end{align}
The condition $K_{\sigma s} = 1$ for $s=\pm$ will also be enforced by $\mathrm{SU}(2)$ invariance \cite{mishmash_continuous_2015}. There are additional marginal terms in the spin sector that will introduce logarithmic corrections to various correlators \cite{schulz_correlation_1990, giamarchi_quantum_2003}, however, we will ignore them since we are not concerned with marginal corrections, but instead with the fixed point the theory flows to.

To understand the scaling dimensions of operators in this theory, we follow the analysis of Ref.~\cite{lai_two-band_2010}. Let $S\in O(2)$ be the matrix such that
\begin{equation}
S^T A S = \begin{pmatrix} A_1& 0\\ 0& A_2\end{pmatrix} \defeq A_D,
\end{equation}
then define $B' = \sqrt{A_D} S^TBS\sqrt{A_D}$. Since $A$ was positive-definite we see that $B'$ is still a real, symmetric, positive definite matrix. Thus, there exists $R\in O(2)$ such that
\begin{equation}
R^TB'R = \begin{pmatrix} B_1'& 0\\ 0& B_2'\end{pmatrix} \defeq B'_D.
\end{equation}
If we now define $\vec{\theta}'_\rho, \vec{\phi}'_\rho$ such that
\begin{align}
\vec{\theta}_\rho =& S\left(\sqrt{A_D}\right)^{-1}R\vec{\theta}'_\rho, \nonumber \\
\vec{\phi}_\rho =& S\sqrt{A_D}R\vec{\phi}'_\rho \label{eqn:prime_coord},
\end{align}
then we note that the primed fields still satisfy Eqs.~(\ref{eqn:comm_relations}) and the action in the charge sector can now be written as
\begin{equation}
\sc{L}^\rho_{\mathrm{C2S2}} = \frac{1}{2\pi}\left[\left|\partial_r \vec{\theta}^{\prime}_\rho\right|^2 + \partial_r \vec{\phi}^{\prime T}_\rho B'_D \partial_r \vec{\phi}'_\rho\right] + \frac{i}{\pi} \partial_r \vec{\theta}'_\rho \cdot \partial_\tau \vec{\phi}'_\rho.
\end{equation}
With this action, it is then straightforward to see that
\begin{equation}
\Delta\left[e^{i\beta\theta^{\prime}_{\rho,i}}\right] = \frac{\beta^2\sqrt{B_i'}}{4}, \text{ and } \Delta\left[e^{i\gamma\phi^{\prime}_{\rho,i}}\right] = \frac{\gamma^2}{4\sqrt{B_i'}}\label{eqn:scaling_dim_prime},
\end{equation}
where $\beta,\gamma\in \bb{R}$, as derived for a theory in this form in Ref.~\cite{giamarchi_quantum_2003}. Using the relations in Eqs.~(\ref{eqn:prime_coord}), the scaling dimension of $e^{i\beta\theta_{\rho\pm}}$ and $e^{i\gamma\phi_{\rho\pm}}$ can then be worked out in terms of the Luttinger parameters in $A$ and $B$. In the spin sector, we must have that
\begin{equation}
\Delta\left[e^{i\beta\theta_{\sigma \pm}}\right] = \frac{\beta^2}{4}, \text{ and } \Delta\left[e^{i\gamma\phi_{\sigma \pm}}\right] = \frac{\gamma^2}{4} \label{eqn:scaling_dim_SU2},
\end{equation}
which is enforced by $\mathrm{SU}(2)$ invariance.

\section{Construction of all possible local operators}

Any local operator must be constructed from the low-energy electrons $c_{Pa\alpha}$ and can thus be expressed as complex linear combinations of powers of them. We will call operators made up solely of powers of the electrons $\sc{O}_p$ for short. Acting with a given $\sc{O}_p$ will then add (or subtract) a well-defined number of particles of a given spin $\alpha$ and parity $P$ to each FS $a$; call this $n_{Pa\alpha}\in \bb{Z}$. Thus, when we bosonize an operator of this type, we can write it as
\begin{align}
\sc{O}_p \sim e^{i \sum_{P,a,\alpha} n_{Pa\alpha}\left(\phi_{a\alpha} + P\theta_{a\alpha}\right)} e^{ir\left(\sum_{P,a,\alpha}n_{Pa\alpha} P q_{Fa}\right)}.
\end{align}
Note that we have neglected possible constants of proportionality, which may include the Klein factors $\eta_{a\alpha}$, as these will not affect the relevance or irrelevance of operators. Additionally, we have ignored terms of the form $c^\dagger_{Pa\alpha} c_{Pa\alpha} = \partial_r(\theta_{a\alpha} + P\phi_{a\alpha})/(2\pi)$. These terms will not create any net particles, i.e., $n_{Pa\alpha} = 0 \ \forall P,a,\alpha$, but will produce additional derivative factors in front of $\sc{O}_p$. However, it is always the case that $\Delta[\partial_r\theta_{a\alpha}] = 1 = \Delta[\partial_r\phi_{a\alpha}]$ \cite{giamarchi_quantum_2003}. Including such terms can thus only decrease the relevance of a given $\sc{O}_p$, and we will therefore ignore them in favor of the bare $\sc{O}_p$.

With this reasoning, we can then write any given local operator as
\begin{align}
\sc{O}(f) =& \sum_{\vec{n}\in \bb{Z}^8} f(\vec{n})\sc{O}_p(\vec{n}) \text{ where } f\colon \bb{Z}^8\rightarrow \bb{C}, \label{eqn:generic_operator}\\
\sc{O}_p(\vec{n}) \defeq& e^{i \sum_{P,a,\alpha} n_{Pa\alpha}\left(\phi_{a\alpha} + P\theta_{a\alpha}\right)} e^{iQ(\vec{n})r} \text{ and}\\
Q(\vec{n}) \defeq& \sum_{P,a,\alpha} n_{Pa\alpha} P q_{Fa} \label{eqn:Q_from_n}.
\end{align}
This then allows for complex linear combinations to be made of operators that add a well-defined number of particles of each species $\vec{n}$. Again, we have neglected possible derivative terms due to their lower relevance.

Ultimately, we want to understand these operators in terms of the basis $\theta_{\rho\pm}, \theta_{\sigma \pm}$ and likewise for $\phi$, as these are the form of the bosonized terms we will be referring to. Using Eqs.~(\ref{eqn:defn_charge_spin}) and (\ref{eqn:defn_total_relative}) we see that
\begin{align}
\theta_{a\alpha} =& \frac{1}{2}(\theta_{\rho+} + \sigma_a \theta_{\rho-} + \sigma_\alpha \theta_{\sigma+} + \sigma_\alpha \sigma_a\theta_{\sigma-}) \text{ and}\\
\phi_{a\alpha} =& \frac{1}{2}(\phi_{\rho+} + \sigma_a \phi_{\rho-} + \sigma_\alpha \phi_{\sigma+} + \sigma_\alpha \sigma_a\phi_{\sigma-}), \text{ where}\\
\sigma_a =& \begin{cases} +1 &\mbox{if } a=1\\ -1 &\mbox{if } a=2\end{cases} \text{ and}\\
\sigma_\alpha =& \begin{cases} +1 &\mbox{if } \alpha=\uparrow\\ -1 &\mbox{if } \alpha=\downarrow\end{cases}.
\end{align}
We can then insert these expressions and rewrite the exponent of $\sc{O}_p(\vec{n})$ in terms of these variables, i.e., it will be given by
\begin{equation}
\sc{O}_p(\vec{n}) \sim \exp\left[i\sum_{\mu = \rho,\sigma;s=\pm}\left(b_{\theta_{\mu s}} \theta_{\mu s} + b_{\phi_{\mu s}} \phi_{\mu s}\right) \right]\nonumber,
\end{equation}
where the coefficients $b$ will be given by
\begin{align}
b_{\theta_{\rho+}} =& \frac{1}{2}\sum_{P,a,\alpha} n_{Pa\alpha} P,\\
b_{\theta_{\rho-}} =& \frac{1}{2}\sum_{P,a,\alpha} n_{Pa\alpha} P\sigma_a,\\
b_{\theta_{\sigma+}} =& \frac{1}{2} \sum_{P,a,\alpha}n_{Pa\alpha} P \sigma_\alpha,\\
b_{\theta_{\sigma-}} =& \frac{1}{2} \sum_{P,a,\alpha}n_{Pa\alpha} P \sigma_\alpha\sigma_a, \\
b_{\phi_{\rho+}} =& \frac{1}{2}\sum_{P,a,\alpha} n_{Pa\alpha}, \label{eqn:b_phi_rho_plus}\\
b_{\phi_{\rho-}} =& \frac{1}{2}\sum_{P,a,\alpha} n_{Pa\alpha} \sigma_a,\\
b_{\phi_{\sigma+}} =& \frac{1}{2} \sum_{P,a,\alpha}n_{Pa\alpha} \sigma_\alpha, \label{eqn:b_phi_sigma_plus}\\
b_{\phi_{\sigma-}} =& \frac{1}{2}\sum_{P,a,\alpha} n_{Pa\alpha}\sigma_\alpha \sigma_a.
\end{align}
It is then straightforward to invert this expression and rewrite the particles created by $\sc{O}_p(\vec{n})$, $\vec{n}$, in terms of the coefficients $b$:
\begin{align}
n_{Pa\alpha} =& \frac{1}{4}Pb_{\theta_{\rho+}} + \frac{1}{4}P\sigma_a b_{\theta_{\rho-}} + \frac{1}{4}P\sigma_\alpha b_{\theta_{\sigma+}}  + \frac{1}{4} P\sigma_\alpha \sigma_a b_{\theta_{\sigma-}}\nonumber \\
&+ \frac{1}{4}b_{\phi_{\rho+}} + \frac{1}{4}\sigma_a b_{\phi_{\rho-}} + \frac{1}{4} \sigma_\alpha b_{\phi_{\sigma+}} + \frac{1}{4} \sigma_\alpha \sigma_a b_{\phi_{\sigma-}} \label{eqn:n_from_b}.
\end{align}
The fact that $n_{Pa\alpha}\in \bb{Z}$ will constrain the possible values that the coefficients $b$ can take.

We can further use this expression for $\vec{n}$ in terms of the coefficients $b$ to express the total momentum $Q$ of the operator $\sc{O}_p(\vec{n})$ in terms of the $b$ coefficients,
\begin{equation}
Q(\{b\}) = (q_{F1}+q_{F2})b_{\theta_{\rho +}} + (q_{F1} - q_{F2})b_{\theta_{\rho-}} \label{eqn:Q_from_b},
\end{equation}
where the sum over $P$ and $\alpha$ annihilated the other terms. This makes it clear that $\theta_{\rho+}$ is the operator associated with the total momentum, while $\theta_{\rho-}$ is the operator associated with the relative momentum. Note that the expression for $Q$ in Eq.~(\ref{eqn:Q_from_n}) shows that $b_{\theta_{\rho+}} + b_{\theta_{\rho-}}$ and $b_{\theta_{\rho+}} - b_{\theta_{\rho-}}$ must be integers. In particular, this implies that $b_{\theta_{\rho+}}$ and $b_{\theta_{\rho-}}$ must be half integer.

\subsection{Possible local CDW operators}
\label{app:CDW_enumerated}

We seek operators which have $b_{\theta_{\rho+}}\neq 0$ with all other coefficients zero. From Eq.~(\ref{eqn:n_from_b}) we see that such an operator will add (or subtract) $n_{Pa\alpha} = Pb_{\theta_{\rho+}}/4$ electrons of spin $\alpha$ and parity $P$ to each FS $a$. Since this must be an integer, we require $b_{\theta_{\rho+}} = 4l$ for some $l\in \bb{Z}$. Then, from Eq.~(\ref{eqn:Q_from_b}), we see that operators of this type will add a total momentum $Q = 4(q_{F1}+q_{F2})l$. From the Luttinger sum rule in Eq.~(\ref{eqn:Luttinger}), we can rewrite this as $Q = 4\pi l\nu_c$. We may therefore write the bosonized form of every operator which depends only on $\theta_{\rho+}$ as 
\begin{equation}
\sc{O}_p(\vec{n}) \sim \sc{O}_{\mathrm{CDW}}^l, \text{ where } \sc{O}_{\mathrm{CDW}} \sim e^{-4i\theta_{\rho+}} e^{-2\pi i (2\nu_c) r}.
\end{equation}
Since $b_{\theta_{\rho+}} = 4$ for $\sc{O}_{\mathrm{CDW}}$, we know that $n_{Pa\alpha}=P$ and we can thus write
\begin{equation}
\sc{O}_{\mathrm{CDW}} = c^\dagger_{R1\uparrow}c^\dagger_{R1\downarrow}c^\dagger_{R2\uparrow}c^\dagger_{R2\downarrow}c_{L1\uparrow}c_{L1\downarrow}c_{L2\uparrow}c_{L2\downarrow}
\end{equation}
in terms of the low-energy fermionic degrees of freedom.

\subsection{Local operators consistent with symmetries of $\sc{L}_{\mathrm{C2S2}}$}
\label{app:operators_consistent}

We enumerate the constraints on operators required by the symmetries described in Sec.~\ref{sec:dominant_umklapp}. Note that some of these constraints were mentioned in Ref.~\cite{sheng_spin_2009}, which treated the case where $\nu_c=1/2$.

Particle number conservation means that the total number of particles an operator $\sc{O}_p(\vec{n})$ creates must be zero, i.e., that $\sum_{P,a,\alpha} n_{Pa\alpha} = 0$. Equation (\ref{eqn:b_phi_rho_plus}) shows that this must mean the coefficient of $\phi_{\rho+}$ is zero, i.e., that $b_{\phi_{\rho+}}=0$. This again agrees with our understanding since $\phi_{\rho+}$ is the conjugate phase to the total density.

Conservation of total spin, $\sum_r S^z_r = \sum_{r,\alpha} \sigma_\alpha c^\dagger_{r\alpha}c_{r\alpha}$, means that $\sum_{P,a,\alpha}n_{Pa\alpha} \sigma_\alpha = 0$, or from Eq.~(\ref{eqn:b_phi_sigma_plus}) $b_{\phi_{\sigma+}} = 0$. This is because $\phi_{\sigma+}$ is the conjugate phase for the total spin.

Conservation of crystal momentum means that $Q = 0 \pmod{2\pi}$, which we see from Eq.~(\ref{eqn:Q_from_b}) will impose constraints on the coefficients of $\theta_{\rho+}$ and $\theta_{\rho-}$. From the Luttinger sum rule, Eq.~(\ref{eqn:Luttinger}), we know that $2m(q_{F1}+q_{F2}) = 2\pi$ and therefore that we may write
\begin{equation}
Q(\{b\}) = 2\pi \frac{b_{\theta_{\rho+}}}{2m} + (q_{F1} - q_{F2}) b_{\theta_{\rho-}} = 0\pmod{2\pi}
\end{equation}
Since $b_{\theta_{\rho+}}$ must be half integer, we thus see that $4mQ = 2\pi (2b_{\theta_{\rho+}}) + 4m(q_{F1}-q_{F2})b_{\theta_{\rho-}} = 4m(q_{F1}-q_{F2})b_{\theta_{\rho-}} = 0 \pmod{2\pi}$. Without fine tuning the momenta, we expect that $q_{F1}-q_{F2}$ will not be some rational multiple of $2\pi$ and thus that there is no nonzero half-integer $b_{\theta_{\rho-}}$ which can be chosen to satisfy this requirement. We may therefore conclude that in the absence of fine tuning conservation of crystal momentum implies that $b_{\theta_{\rho-}} = 0$. This then means that $Q = 2\pi b_{\theta_{\rho+}}/2m = 0 \pmod{2\pi}$ and thus that $b_{\theta_{\rho+}} \in 2m\bb{Z}$.

Finally, we must address: time-reversal invariance, the point group symmetries of the 1D chain, and $\mathrm{SU}(2)$ invariance. Unlike the other symmetries, these will not constrain the terms allowed in the exponential of a generic operator, however, we will find that they constrain the ways that operators must be added together. In other words, these symmetries will not act to constrain the $b$-coefficients but rather the $f\colon \bb{Z}^8 \rightarrow \bb{C}$ that defines the coefficients of each $\sc{O}_p(\vec{n})$. We consider a combination of time reversal and a spin rotation which we label $\sc{T}$,
\begin{align}
\sc{T}\colon& c_{Pa\alpha}(r) \mapsto \sc{K}c_{-Pa\alpha}(r)\\
\implies& \theta_{a\alpha}(r) \mapsto \theta_{a\alpha}(r)\\
& \phi_{a\alpha}(r) \mapsto -\phi_{a\alpha}(r),
\end{align}
where $\sc{K}$ is the conjugation operator. Time reversal alone would have taken $c_{Pa\alpha}\rightarrow \sc{K} (i\sigma^y)_{\alpha \beta} c_{-Pa\beta}$ so the spin operator was flipped, but this would have been more complicated in bosonized form. Since $\mathrm{SU}(2)$ is also a good symmetry of the system, we have considered $\sc{T}$ for ease. Note that this preserves $\rho_{a\alpha}(r) \mapsto \rho_{a\alpha}(r)$ and the canonical commutation relation $[\rho_{a\alpha}(r), \phi_{a\alpha}(r')] = i\delta(r-r')$ as it must. Indeed, we could have used the preservation of these two to deduce the action of $\sc{T}$ on $\theta_{a\alpha}$ and $\phi_{a\alpha}$. Next, the only point-group symmetry of the lattice pictured in Fig.~\ref{fig:setup}(a) is the symmetry $\sigma_{\mathrm{site}} \colon r \mapsto -r$ which reflects about a given point. Since $\sigma_{\mathrm{site}}$ flips spatial directions, but not time, it must also take $P\mapsto -P$. Thus,
\begin{align}
\sigma_{\mathrm{site}} \colon& c_{Pa\alpha}(r) \mapsto c_{-Pa\alpha}(-r)\\
\implies& \theta_{a\alpha}(r) \mapsto -\theta_{a\alpha}(-r)\\
& \phi_{a\alpha}(r) \mapsto \phi_{a\alpha}(-r).
\end{align}
We see that this symmetry will take the densities $\rho_{a\alpha}(r) = \partial_r \theta_{a\alpha}/\pi \mapsto \rho_{a\alpha}(-r)$ as it must.

Finally, we must consider $\mathrm{SU}(2)$ invariance. This is more complicated to express in terms of bosonized operators. However, none of these mappings will change the scaling of the resulting operator \cite{giamarchi_quantum_2003}.

Returning to Eq.~(\ref{eqn:n_from_b}), we finally see this means that
\begin{align}
n_{Pa\alpha} =& \frac{1}{2}Pml + \frac{1}{4} P\sigma_\alpha b_{\theta_{\sigma+}} + \frac{1}{4}P\sigma_\alpha \sigma_a b_{\theta_{\sigma-}} \nonumber \\
&+ \frac{1}{4}\sigma_a b_{\phi_{\rho-}} + \frac{1}{4}\sigma_\alpha \sigma_a b_{\phi_{\sigma-}} \label{eqn:generic_nval}
\end{align}
describes the possible ways we can add particles consistent with all of these symmetries. We also see that operators which preserve these symmetries will scale as
\begin{align}
\Delta[\sc{O}_p(\vec{n})] =& (ml)^2K_{\theta_{\rho+}} + \frac{1}{4}b_{\phi_{\rho-}}^2K_{\phi_{\rho-}} \nonumber \\
&+ \frac{1}{4} \left(b_{\theta_{\sigma+}}^2 + b_{\theta_{\sigma-}}^2 + b_{\phi_{\sigma-}}^2 \right) \label{eqn:generic_scaling},
\end{align}
where we are again defining $K_{\phi_{\rho-}}$ to be the scaling dimension of $e^{2i\phi_{\rho-}}$. Note that since $A$ and $B$ may not be diagonal, it is not generally true that $K_{\phi_{\rho-}}$ is the inverse of $K_{\theta_{\rho-}}$, where this is likewise defined to be the scaling dimension of $e^{2i\theta_{\rho-}}$.

\subsection{Most relevant operators consistent with symmetries of $\sc{L}_{\mathrm{C2S2}}$}
\label{app:most_relevant_operators}

The operators which might possibly drive a transition out of the C2S2 metallic state are those which are most relevant, so we will enumerate the operators consistent with the constraint Eq.~(\ref{eqn:generic_nval}) that have the smallest scaling dimension given by Eq.~(\ref{eqn:generic_scaling}). Consider the integers defined by Eq.~(\ref{eqn:generic_nval}) and sum over $\sigma_\alpha$. Then it is clear that $\sigma_a b_{\phi_{\rho-}}/2 \in \bb{Z}$ and thus that $b_{\phi_{\rho-}} = 2l_{\phi_{\rho-}}$ for some $l_{\phi_{\rho-}}\in \bb{Z}$. If we similarly sum over $\sigma_a$, we can see that $b_{\theta_{\sigma+}} = 2l_{\theta_{\sigma+}}$ for $l_{\theta_{\sigma+}}\in \bb{Z}$. Then summing over $P$ and using the fact that $b_{\phi_{\rho-}} = 2l_{\phi_{\rho-}}$, we can see that $b_{\phi_{\sigma-}} = 2l_{\phi_{\sigma-}}$ for $l_{\phi_{\sigma-}}\in \bb{Z}$. Finally, by multiplying by $P$, summing over it, and using the fact that $b_{\theta_{\sigma+}} = 2l_{\theta_{\sigma+}}$, we may conclude that $b_{\theta_{\sigma-}} = 2l_{\theta_{\sigma-}}$ with $l_{\theta_{\sigma-}}\in \bb{Z}$. We therefore have that the number of particles created by these local operators must be of the form
\begin{align}
n_{Pa\alpha} =& \frac{1}{2}\left(Pml + P\sigma_\alpha l_{\theta_{\sigma+}} + P\sigma_\alpha \sigma_a l_{\theta_{\sigma-}}\right. \nonumber\\
&\left.+ \sigma_a l_{\phi_{\rho-}} + \sigma_\alpha \sigma_a l_{\phi_{\sigma-}}\right)  \in \bb{Z}\label{eqn:generic_nval_int},
\end{align}
and their scaling dimensions will be given by
\begin{align}
\Delta[\sc{O}_p(\vec{n})] =& (ml)^2 K_{\theta_{\rho+}} + l^2_{\phi_{\rho-}} K_{\phi_{\rho-}}\nonumber \\
&+ l^2_{\theta_{\sigma+}} + l^2_{\theta_{\sigma-}} + l^2_{\phi_{\sigma-}} \label{eqn:generic_scaling_int}.
\end{align}

Operators involving just the spins will be at worst marginal and because of their $\mathrm{SU}(2)$ invariance will be unaffected by tuning the theory towards an insulator. We will therefore ignore such options, considering instead operators that include at least one of $\theta_{\rho+}$ or $\phi_{\rho-}$, as these are the terms that can possibly tune us across a critical point by tuning $K_{\theta_{\rho+}}$ and $K_{\phi_{\rho-}}$.

\subsubsection{Even $m$}

For the case of even $m$, the $Pml/2$ term in Eq.~(\ref{eqn:generic_nval_int}) will be an integer no matter the value of $l$, so the constraint of Eq.~(\ref{eqn:generic_nval_int}) reduces to 
\begin{align}
P\sigma_\alpha l_{\theta_{\sigma+}} + P\sigma_\alpha \sigma_a l_{\theta_{\sigma-}} + \sigma_a l_{\phi_{\rho-}} + \sigma_\alpha \sigma_a l_{\phi_{\rho-}}
\end{align}
being an even integer.

Now we will find the most relevant operator subject to this constraint which does not involve the total charge mode $\theta_{\rho+}$ but does involve $\phi_{\rho-}$, i.e., $l=0$ and $l_{\phi_{\rho-}} \neq 0$. Clearly, the smallest we can then make Eq.~(\ref{eqn:generic_scaling_int}) is $l_{\phi_{\rho-}} = 1$ with all other coefficients zero. However, this does not satisfy the constraint above. The next most relevant operators will then either have $l_{\phi_{\rho-}}=2$ with all other coefficients zero, or $l_{\phi_{\rho-}} = 1$ with one of the coefficients of the spin operators as small as possible without being zero, say $l_{\phi_{\sigma-}} = 1$. Both of these will satisfy the even integer constraint. The case of $l_{\phi_{\rho-}} = 2$ will have $n_{Pa\alpha} = \sigma_a$ with this operator described by
\begin{equation}
\sc{O}_p(\vec{n}) \sim c^\dagger_{R1\uparrow}c^\dagger_{R1\downarrow}c^\dagger_{L1\uparrow}c^\dagger_{L1\downarrow}c_{R2\uparrow}c_{R2\downarrow}c_{L2\uparrow}c_{L2\downarrow} \sim e^{-4i\phi_{\rho-}}.
\end{equation}
The operator $\cos(4\phi_{\rho-})$, which has the same scaling dimension, will be consistent with all of the symmetries enumerated in the previous section. It will have a scaling dimension of $\Delta[\sc{O}_p(\vec{n})] = 4K_{\phi_{\rho-}}$. Next consider the case of $l_{\phi_{\rho-}} = 1$ and $l_{\phi_{\sigma-}} = 1$. From Eq.~(\ref{eqn:generic_nval}), we see that
\begin{align}
n_{Pa\alpha} =& \frac{\sigma_a}{2}(1+\sigma_\alpha) \in \bb{Z}.
\end{align}
Thus, we may write this operator as
\begin{align}
\sc{O}_p(\vec{n}) \sim& c^\dagger_{R1\uparrow}c^\dagger_{L1\uparrow}c_{R2\uparrow}c_{L2\uparrow}\\
\sim& e^{-2i(\phi_{\rho-} + \phi_{\sigma-})}.
\end{align}
It can be checked that the operator $\cos(2\phi_{\rho-})\cos(2\phi_{\sigma-})$, which has the same scaling dimension, is consistent with all symmetries of the system, provided the necessary spin terms are added so $\mathrm{SU}(2)$ invariance is respected. This operator will have scaling dimensions of $\Delta[\sc{O}_p(\vec{n})] = K_{\phi_{\rho-}} + 1$. We now claim that $\cos(2\phi_{\rho-})\cos(2\phi_{\sigma-})$ is always a more relevant operator than $\cos(4\phi_{\rho-})$ on the metallic side and at the critical point. Suppose it were not, then $4K_{\phi_{\rho-}} \leq  K_{\phi_{\rho-}} + 1$ and $K_{\phi_{\rho-}} \leq 1/3$. But this would mean that $K_{\phi_{\rho-}} + 1 \leq 4/3 <2$, so the system would have already gapped out the relative charge mode. Thus, the most relevant operator which involves $\phi_{\rho-}$ but not $\theta_{\rho+}$ is $\cos(2\phi_{\rho-})\cos(2\phi_{\sigma-})$ plus any additional operators required by $\mathrm{SU}(2)$ invariance, which will have the same scaling dimension. 

We now consider the most relevant operator involving $\theta_{\rho+}$ but not $\phi_{\rho-}$, i.e., $l\neq 0$ but $l_{\phi_{\rho-}} = 0$. Clearly, the smallest we can then make Eq.~(\ref{eqn:generic_scaling_int}) is $m^2K_{\theta_{\rho+}}$. This operator will have $n_{Pa\alpha} = Pml/2$, which is indeed an integer since $m$ is even. Indeed, this just describes the umklapp operator, $\cos(2m\theta_{\rho+})$. Thus, the umklapp operator is the most relevant operator involving $\theta_{\rho+}$ but not $\phi_{\rho-}$.

Finally, we consider the most relevant operator involving both $\theta_{\rho+}$ and $\phi_{\rho-}$. As mentioned, because $m$ is even, the $Pml/2$ term dropped out of the integer constraint on the coefficients. This means that we are free to choose $l=1$ without affecting the constraints on the other coefficients. Then, by the same logic as when we considered the most relevant operator involving $\phi_{\rho-}$ and not $\theta_{\rho+}$, the most relevant operators which involve both must scale like $m^2 K_{\theta_{\rho+}} + K_{\phi_{\rho-}} + 1$ or $m^2 K_{\theta_{\rho+}} + 4K_{\phi_{\rho-}}$. Again, by the same logic as above, the operator scaling like $m^2 K_{\theta_{\rho+}} + K_{\phi_{\rho-}} + 1$ must be the more relevant of the two. But this operator is obviously less relevant than both the umklapp operator and the operator $\cos(2\phi_{\rho-})\cos(2\phi_{\sigma-})$ since it scales as their product. We may thus neglect it.

We conclude by noting that for even $m$ the most relevant operators which might tune us out of the C2S2 metallic phase are the umklapp operator, $\cos(2m\theta_{\rho+})$, and the operator $\cos(2\phi_{\rho-})\cos(2\phi_{\sigma-})$. In particular, we note that if we wanted to tune to any phase which gaps out $\phi_{\rho-}$ we must gap out both spins. Thus, attempting to gap out both charge modes upon exiting the C2S2 metal must necessarily gap out both spin modes as well.

\subsubsection{Odd $m$}

The case of odd $m$ will be different. Now Eq.~(\ref{eqn:generic_nval_int}) reduces to the condition that 
\begin{align}
P\sigma_\alpha l_{\theta_{\sigma+}} + P\sigma_\alpha \sigma_a l_{\theta_{\sigma-}} + \sigma_a l_{\phi_{\rho-}} + \sigma_\alpha \sigma_a l_{\phi_{\rho-}}
\end{align}
be an even integer when $l$ is even and an odd integer when $l$ is odd.

Let us first consider the most relevant operator subject to this constraint which does not involve the total charge mode but does involve $\phi_{\rho-}$, i.e., $l=0$ and $l_{\phi_{\rho-}}\neq 0$. Since $l=0$ is even, the above constraint will be identical to the case of $m$ even, and we conclude that the most relevant operator in this case must again be $\cos(2\phi_{\rho-})\cos(2\phi_{\sigma-})$.

We next consider the most relevant operator subject to this constraint which involves $\theta_{\rho+}$ but not $\phi_{\rho-}$. As in the case of even $m$, the smallest we can make Eq.~(\ref{eqn:generic_scaling_int}) is $m^2K_{\theta_{\rho+}}$, when $l=1$ and all other coefficients are zero. But this would then imply that $n_{Pa\alpha} = Pml/2 \notin \bb{Z}$ since $m$ is odd. We must then either take $l=2$ with all other coefficients being equal to zero or consider $l=1$ with one of the coefficients of the spin operators as small as possible, say $l_{\theta_{\sigma+}} = 1$. The case of $l=2$ will have $n_{Pa\alpha} = Pm$, which just describes the operator $\sc{O}_{\mathrm{CDW}}$ raised to the $m$th power. This operator is clearly consistent with all possible symmetries as it can just be thought of as the umklapp operator squared, and will scale as $4m^2K_{\theta_{\rho+}}$. The case $l=1$ with $l_{\theta_{\sigma+}} = 1$ just describes the umklapp operator when $m$ is odd. This operator is clearly consistent with all possible symmetries and will scale as $m^2K_{\theta_{\rho+}} + 1$. The umklapp operator, $\cos(2m\theta_{\rho+})\cos(2\theta_{\sigma+})$, will be the more relevant of the two on the metallic side and at the critical point, as if it was not then $K_{\theta_{\rho+}} \leq 1/(3m^2)$ and $m^2K_{\theta_{\rho+}} + 1 \leq 4/3 < 2$.

Finally, we consider the most relevant object which involves both $\theta_{\rho+}$ and $\phi_{\rho-}$. The smallest we can make Eq.~(\ref{eqn:generic_scaling_int}) is then $m^2K_{\theta_{\rho+}} + K_{\phi_{\rho-}}$ by choosing $l = 1$ and $l_{\phi_{\rho-}}=1$. For this operator, $n_{Pa\alpha} = (Pm + \sigma_a)/2$, which is indeed an integer. One can check that $\cos(2m\theta_{\rho+})\cos(2\phi_{\rho-})$ is also consistent with all possible symmetries. Thus the most relevant operator containing both $\theta_{\rho+}$ and $\phi_{\rho-}$ is $\cos(2m\theta_{\rho+})\cos(2\phi_{\rho-})$.

The three most relevant possible operators which might tune us out of the C2S2 metallic phase for $m$ odd are the umklapp operator, $\cos(2m\theta_{\rho+})\cos(2\theta_{\sigma+})$, and the operators $\cos(2\phi_{\rho-})\cos(2\phi_{\sigma-})$ and $\cos(2m\theta_{\rho+})\cos(2\phi_{\rho-})$. We now claim that $\cos(2m\theta_{\rho+})\cos(2\phi_{\rho-})$ must be of lesser or the same relevance than the other two operators. Suppose it was not, then $m^2K_{\theta_{\rho+}} + K_{\phi_{\rho-}}$ would be less than either $K_{\phi_{\rho-}}+1$ or $m^2K_{\theta_{\rho+}} + 1$. In the first case, we would have $m^2K_{\theta_{\rho+}} < 1$ and thus $m^2K_{\theta_{\rho+}} + 1< 2$, so the umklapp operator would have driven a transition out of the metal. In the second case, we would have that $K_{\phi_{\rho-}} < 1$, so the $\cos(2\phi_{\rho-})\cos(2\phi_{\sigma-})$ operator would have already driven a transition out of the metal. We thus conclude that $m^2 K_{\theta_{\rho+}} + K_{\phi_{\rho-}} \geq m^2K_{\theta_{\rho+}} +1, K_{\phi_{\rho-}} +1$, with equality only possible at the critical point. This means that the two most relevant operators which can tune us out of the C2S2 metal both have spin operators present. Thus, tuning out of the C2S2 metal must always gap out the spin for $m$ odd even if only one charge mode is gapped out. The intuition for this in the case of the total charge is discussed in the main text.

\onecolumngrid
\section{Physical properties of WM transition.}

\subsection{Even $m$ (C2S2 $\rightarrow$ C1S2)}
\label{app:evenm_correlations}

We wish to find the logarithmic corrections to $\langle \sc{O}_{\mathrm{CDW}}(r)\sc{O}_{\mathrm{CDW}}(0)\rangle$ in the theory with $\sc{L}^\rho_{\mathrm{CDW}}$ diagonal in the $\rho\pm$ basis. Note that in this theory $K_{\theta_{\rho+}}$ will become a true Luttinger parameter, so we will write $K_{\rho+}$ instead. We will also write $v=v_{\rho+}$ for short. We now restate some basics. For the Lagrangian 
\begin{equation}
\sc{L}^0 = \sc{L}^\sigma_{\mathrm{C2S2}} + \frac{1}{2\pi}\sum_{s=\pm}\left[\frac{1}{v_{\rho s}}(\partial_\tau \theta_{\rho s})^2 + v_{\rho s}(\partial_r \theta_{\rho s})^2 \right],
\end{equation}
we have the correlation function,
\begin{equation}
\langle [\theta_{\rho+}(r,\tau) - \theta_{\rho+}(0)]^2\rangle_0 = F(r,\tau) \equiv \frac{1}{2}\ln\left[\frac{x^2 + (v_{\rho+}|\tau|+\alpha)^2}{\alpha^2}\right],
\end{equation}
where $\alpha$ is a real-space cut-off that is taken to zero. The details of this cutoff are in Appendix C of Ref.~\cite{giamarchi_quantum_2003}. We will make the definition that $\tilde{r} = (r, v\tau)$ to take advantage of the Lorentz invariance of the $\theta_{\rho+}$ part of the action. With this definition $F(r,\tau) \sim \ln(\tilde{r}/\alpha)$ in the large $\tilde{r}/\alpha$ limit.

Using these results, we can now compute the RG equations by studying the real space renormalization of correlation functions. We will have that
\begin{align}
R_a(\tilde{r}) \equiv& \left\langle e^{ia\theta_{\rho+}(\tilde{r})} e^{-ia\theta_{\rho+}(0)}\right\rangle\\
=& R_a(\tilde{r})^{(0)} + \frac{1}{2}g_u^2R_a(\tilde{r})^{(2)} + \sc{O}(g_u^4),
\end{align}
where we can expand perturbatively in $g_u$; here $g_u$ is the coefficient of the umklapp term as written in the full theory in Eq.~(\ref{eqn:even_m_full_theory}). The odd order terms are zero because the coefficients of $\theta_{\rho+}$ in the exponentials cannot possibly sum to zero. We can then evaluate this term by term. The expansion is exactly analogous to the treatment of the KT transition in Chap. 2 of Ref.~\cite{giamarchi_quantum_2003}, so we simply quote the result here. To second order, we will have that
\begin{equation}
R_a(\tilde{r}) = e^{-a^2 K_{\rho+}F(\tilde{r})/2}\left[1 + \frac{a^2}{2}\frac{g_u^2}{4\pi^2 v^2}m^2K_{\rho+}^2 F(\tilde{r}) \int_{\tilde{r}_s>\alpha} \frac{\td \tilde{r}_s}{\alpha} \ \left(\frac{\tilde{r}_s}{\alpha}\right)^{3-2m^2K_{\rho+}}\right].
\end{equation}
This looks like the zeroth order correlation function, but with an effective
\begin{equation}
K^*_{\rho+}(\alpha) = K_{\rho+}(\alpha) - \frac{g_u^2}{4\pi^2v^2}(\alpha) m^2K_{\rho+}^2(\alpha) \int_{\tilde{r}_s>\alpha} \frac{\td \tilde{r}_s}{\alpha} \ \left(\frac{\tilde{r}_s}{\alpha}\right)^{3-2m^2K_{\rho+}}.
\end{equation}
We now imagine letting the real-space cutoff flow to an infinitesimally larger value, i.e., taking $\alpha' = \alpha(1+\td l)$. The effective value of $K_{\rho+}$ should be unchanged under this procedure, since the long-distance physics should be unaffected by our choice of a cutoff. Setting $K^*_{\rho+}(\alpha') = K^*_{\rho+}(\alpha)$ immediately gives us the RG equations:
\begin{align}
\frac{\td K_{\rho+}}{\td l} =& - y^2 m^2K_{\rho+}^2,\\
\frac{\td y}{\td l} =& (2 - m^2K_{\rho+})y,
\end{align}
where $y = g_u/2\pi v$. If we linearize about the critical point $K_{\rho+} = 2/m^2$ by taking $x = 2-m^2K_{\rho+}$, then we see to lowest order our differential equations are
\begin{align}
\frac{\td x}{\td l} =& 4y^2,\\
\frac{\td y}{\td l} =& xy.
\end{align}
The trajectories are thus the standard hyperbolas of the KT-transition. In the disordered regime, where $2|y|>|x|$, they can be solved via
\begin{align}
x(l) =& A\tan\left[Al + \tan^{-1}\left(\frac{x_0}{A}\right)\right],\\
y(l) =& \frac{A}{2}\sec\left[Al + \tan^{-1}\left(\frac{x_0}{A}\right)\right],
\end{align}
where $A = \sqrt{4y^2 - x^2}$ is a constant of the flow. If we now take the flow close to the critical point, in the limit where $x_0/A, Al\ll 1$, then we see that the trajectory becomes
\begin{align}
x(l) =& \frac{x_0}{1-x_0l},\\
y(l) =& \frac{1}{2}\frac{x_0}{1-x_0l},
\end{align}
which is exactly the form the flow takes on the whole separatrix $2y(l) = x(l)$. Thus, close enough to the critical point the RG flow is identical to the flow on the separatrix.

We have derived the renormalization equations by computing a specific correlator and demanding that it be unchanged under the RG flow. This was done perturbatively in the umklapp coupling. However, we have not addressed a deeper problem. We note that the $g_u^2$ correction to $R_a(\tilde{r})^{(0)}$ has a factor of $F(\tilde{r}) \sim \ln(|\tilde{r}|/\alpha)$, which is divergent as $|\tilde{r}|/\alpha \rightarrow \infty$. We should not trust this divergence, however, as it would go away for a fixed but large $|\tilde{r}|$ if we chose $\alpha \sim |\tilde{r}|$. Thus, even though we are choosing $K_{\theta_{\rho+}}(l)$ and $g_u(l)$ such that the effective $g_u$ and $K_{\theta_{\rho+}}$ do not change under the RG flow, the presence of this logarithm in the correlator means that the RG flow needs to be treated in the correlator as well. Note that this is a generic feature of any correlator with a marginally irrelevant operator in the theory, as we can expect logarithms that diverge at long distances to appear in these theories. To treat this, we write $R_a(\tilde{r},\alpha) = I_a(\alpha,\alpha')R_a(\tilde{r},\alpha')$, which allows us to treat the RG flow of the correlator by incorporating it into the $I_a$ function. Dividing out by a factor of $R_a(\tilde{r})^{(0)}$, $\overline{R}_a(\tilde{r}, \alpha) \equiv R_a(\tilde{r},\alpha)/R_a(\tilde{r})^{(0)}$, allows us to consider an $I_a$ which is $1$ in the $g_u=0$ limit and thus treat it perturbatively. If we now take $\alpha' = |\tilde{r}|$, then we can expect that $\overline{R}_a(\tilde{r}, \alpha') = \sc{O}(1)$ since the $g_u^2F(\tilde{r})$ term will be small in this limit. Thus, we should expect $R_a(\tilde{r},\alpha)/R_a(\tilde{r})^{(0)} \sim I_a(\alpha, |\tilde{r}|)$. To compute $I_a(\alpha, |\tilde{r}|)$, we first find $I_a(\alpha, \alpha e^{\td l})$ and integrate it. We see that
\begin{align}
\overline{R}_a(\tilde{r},\alpha) =& 1 + \frac{a^2}{2}y^2(\alpha) m^2K^2_{\theta_{\rho+}}(\alpha) F(\tilde{r}) \int_{\tilde{r}_s>\alpha} \frac{\td \tilde{r}_s}{\alpha} \left(\frac{\tilde{r}_s}{\alpha}\right)^{3-2m^2K_{\rho+}}\\
=& 1 + \frac{a^2}{2}y^2(\alpha)m^2 K^2_{\rho+}(\alpha) F(\tilde{r})\td l\\
&+ \frac{a^2}{2}y^2(\alpha')m^2 K^2_{\rho+}(\alpha') F'(\tilde{r}) \int_{\tilde{r}_s>\alpha' }\frac{\td \tilde{r}_s}{\alpha'} \ \left(\frac{\tilde{r}_s}{\alpha'}\right)^{3-2m^2K_{\rho+}},
\end{align}
where we expanded in terms of $\alpha' = \alpha e^{\td l}$, used the fact that $y^2(\alpha') = y^2(\alpha) (\alpha'/\alpha)^{4-2m^2K_{\rho+}}$ and $K^2_{\rho+}(\alpha') = K_{\rho+}(\alpha) + \sc{O}(y^2)$ to make this perturbative connection. Thus,
\begin{align}
I_a(\alpha,\alpha e^{\td l}) =& 1 + \frac{a^2}{2}y^2(\alpha)m^2 K^2_{\rho+}(\alpha) F(\tilde{r})\td l\\
\implies \ln\left(I_a(\alpha,\alpha e^{\td l})\right) =&  \frac{a^2}{2}y^2(\alpha)m^2 K^2_{\rho+}(\alpha) F(\tilde{r})\td l.
\end{align}
The definition of $I_a(\alpha,\alpha')$ makes it clear that $I_a(\alpha,\alpha'') = I_a(\alpha, \alpha')I_a(\alpha',\alpha'')$ and thus that
\begin{align}
I_a(\alpha, \tilde{r}) =& \prod_j I_a(\alpha e^{j\td l}, \alpha e^{j\td l}e^{\td l})\\
=& \exp\left[\sum_j \ln\left(I_a(\alpha e^{j\td l}, \alpha e^{j\td l}e^{\td l})\right)\right]\\
=& \exp\left[\frac{a^2}{2}\int_{0}^{\ln(\tilde{r}/\alpha)} \td l \ m^2y^2(l)K^2_{\rho+}(l)\left(\ln(\tilde{r}/\alpha) - l\right) \right],
\end{align}
where we used the fact that $\ln(\tilde{r}/\alpha(l)) = \ln(\tilde{r}/\alpha e^l) = \ln(\tilde{r}/\alpha) - l$. From the RG equations, we note that $m^2 y^2(l) K^2_{\rho+}(l) = - \td K_{\rho+}/\td l$ and thus
\begin{align}
I_a(\alpha,\tilde{r}) =& \exp\left[-\frac{a^2}{2}\int^{\ln(\tilde{r}/\alpha)}_0 \td l \ \frac{\td K_{\rho+}}{\td l}(\ln(\tilde{r}/\alpha) - l)\right]\\
=& \exp\left[\frac{a^2}{2}K_{\rho+}\ln\left(\frac{\tilde{r}}{\alpha}\right) - \frac{a^2}{2}\int_0^{\ln(\tilde{r}/\alpha)} \td l \ K_{\rho+}(l)\right]\\
\implies R_a(\tilde{r},\alpha) \sim& R_a(\tilde{r},\alpha)^{(0)}I_a(\alpha,\tilde{r})\\
=& \exp\left[-\frac{a^2}{2}\int_0^{\ln(\tilde{r}/\alpha)}\td l \ K_{\rho+}(l)\right].
\end{align}

We have thus learned that the value of $\langle e^{ia\theta_{\rho+}(r)}e^{ia\theta_{\rho+}(0)}\rangle$ requires tracking the full RG flow of $K_{\rho+}(l)$ to avoid the inconsistencies arising from the log divergence of the perturbative corrections. We saw above that close enough to the critical point the RG flow is given by $x(l) = x_0/(1-x_0l)$ for $x = 2-m^2K_{\rho+}$. If we now insert this form, we will find that
\begin{align}
R_a(\tilde{r},\alpha) =& \left(\frac{\alpha}{\tilde{r}}\right)^{a^2/m^2}\exp\left[\frac{a^2}{2m^2} \int_0^{\ln(\tilde{r}/\alpha)} \td l \ \frac{x_0}{1-x_0l}\right]\\
=& \left(\frac{\alpha}{\tilde{r}}\right)^{a^2/m^2}\left(\frac{x(\ln(\tilde{r}/\alpha))}{x_0}\right)^{a^2/2m^2}.
\end{align}
The logarithmic corrections to this correlator are thus particularly simple along the separatrix.

Based on our general arguments in the main text, we should expect that
\begin{align}
\langle \sc{O}_{\mathrm{CDW}}\rangle \sim& \sqrt{\langle \sc{O}_{\mathrm{CDW}}(\xi)\sc{O}_{\mathrm{CDW}}(0)\rangle}\\
=& \sqrt{R_4(\xi,\alpha)}\\
=& \left(\frac{\alpha}{\xi}\right)^{8/m^2}\left(\frac{x(\ln(\xi/\alpha))}{x_0}\right)^{4/m^2}\\
\sim& \exp\left(-\frac{8C/m^2}{\sqrt{2-m^2K_{\theta_{\rho+}}}}\right)(2-m^2K_{\theta_{\rho+}})^{-4/m^2},
\end{align}
since $\xi$ is \textit{defined} by the requirement that $x(l^*)\sim 1$ for $\xi \sim \alpha e^{l^*}$.

\subsection{Odd $m$ (C2S2 $\rightarrow $ C1S0)}
\label{app:oddm_correlations}

The unperturbed theory which is diagonal in the $\rho\pm$ basis and has identical spin and charge velocities can be written as:
\begin{equation}
\sc{L}_{\mathrm{C2S2}} = \sum_{\substack{\nu = \rho,\sigma \\ s = \pm}}\frac{v}{2\pi}\left[\frac{1}{K_{\nu s}}\left(\partial_r \theta_{\nu s}\right)^2 + K_{\nu s} \left(\partial_r \phi_{\nu s}\right)^2\right] + \frac{i}{\pi} (\partial_r \theta_{\nu s})(\partial_\tau \phi_{\nu s}).
\end{equation}
Note that we have not set $K_{\sigma \pm} = 1$ here, even though this will be the value they take in the $\mathrm{SU}(2)$ invariant C2S2 theory. We want to be able to track the evolution into the spin insulator, so we need to allow their values to flow, though we will set their initial values under the flow to be one. With this unperturbed theory, we can write Wick's theorem as
\begin{equation}
\left\langle \prod_j e^{i\left(A_j \theta_{\nu s}(\tilde{r}_j) + B_j\phi_{\nu s}(\tilde{r}_j)\right)}\right\rangle = \exp\left[-\frac{1}{2}\sum_{i<j}\left(-A_iA_j K_{\nu s} - B_i B_j K^{-1}_{\nu s}\right)F(\tilde{r}_i-\tilde{r}_j) + \left(A_iB_j + B_iA_j\right)D(\tilde{r}_i-\tilde{r}_j)\right],
\end{equation}
where, again,
\begin{equation}
F(\tilde{r}) = \frac{1}{2}\ln\left[\frac{r^2 + (v|\tau|+\alpha)^2}{\alpha^2}\right] \ \text{ and } \ D(\tilde{r}) = -i\mathrm{arg}\left[v\tau + \alpha\mathrm{sgn}(\tau) + ir\right].
\end{equation}
Note that $F$ and $D$ are the real and imaginary parts of the function $\ln(v\tau + \alpha\mathrm{sgn}(\tau) - ir)$, respectively. As before, the correlator above will be zero unless $\sum_i A_i = 0 = \sum_i B_i$. We have taken these results directly from Giamarchi \cite{giamarchi_quantum_2003}. We further note that because $\sc{L}_{\mathrm{C2S2}}$ is diagonal in the $\nu s$, modes any expectation values in the unperturbed theory, $\langle \cdot \rangle_0$, will factorize into the different $\nu s$ sectors. 

We can now begin to compute correlators. We define
\begin{equation}
R^{\nu s}_a(r) \equiv \left\langle e^{ia\theta_{\nu s}(\tilde{r})} e^{-ia\theta_{\nu s}(0)}\right\rangle.
\end{equation}
We will then have that
\begin{align}
R^{\rho +}_a(\tilde{r}) =& e^{-a^2 K_{\rho +}F(\tilde{r})/2} + \frac{1}{2}g_1^2 R^{\rho +}_a(\tilde{r})_{g_1}^{(2)} + \frac{1}{2}g_2^2 R^{\rho +}_a(\tilde{r})_{g_2}^{(2)}, \\
R^{\rho +}_a(\tilde{r})^{(2)}_{g_1} =& \sum_{\sigma_1, \sigma_2}\int \frac{\td^2 \tilde{r}' \td^2 \tilde{r}''}{(2\pi \alpha)^4 v^2}\left[\left\langle e^{ia\theta_{\rho +}(\tilde{r})} e^{-ia\theta_{\rho +}(0)} e^{2mi\sigma_1 \theta_{\rho+}(\tilde{r}')} e^{-2mi \sigma_1 \theta_{\rho+}(\tilde{r}'')}\right\rangle_0 \left\langle e^{2i\sigma_2 \theta_{\sigma+}(\tilde{r}')} e^{-2i\sigma_2 \theta_{\sigma+}(\tilde{r}'')}\right\rangle_0  \right.\\
&\left.- \left\langle e^{ia\theta_{\rho +}(\tilde{r})} e^{-ia\theta_{\rho +}(0)}\right\rangle_0\left\langle e^{2mi\sigma_1 \theta_{\rho+}(\tilde{r}')} e^{-2mi \sigma_1 \theta_{\rho+}(\tilde{r}'')}\right\rangle_0 \left\langle e^{2i\sigma_2 \theta_{\sigma+}(\tilde{r}')} e^{-2i\sigma_2 \theta_{\sigma+}(\tilde{r}'')}\right\rangle_0\right]\\
=& 2 e^{-a^2 K_{\rho+}F(\tilde{r})/2} \sum_{\sigma_1}\int \frac{\td^2 \tilde{r}' \td^2 \tilde{r}''}{(2\pi \alpha)^4 v^2} e^{-2(m^2 K_{\rho+} + K_{\sigma+})F(\tilde{r}'-\tilde{r}'')}\left[e^{ma\sigma_1 K_{\rho +}[F(\tilde{r}-\tilde{r}') - F(\tilde{r}-\tilde{r}'') + F(\tilde{r}'') - F(\tilde{r}')]} - 1\right]\\
=& 2m^2a^2 K_{\rho+}^2 e^{-a^2 K_{\rho+}F(\tilde{r})/2} \int \frac{\td^2 \tilde{r}_s \td^2 R}{(2\pi \alpha)^4 v^2} e^{-2(m^2 K_{\rho+} + K_{\sigma+})F(\tilde{r}_s)} \left[\tilde{r}_s\cdot \nabla_R\left(F(\tilde{r}-R) - F(R)\right)\right]^2,
\end{align}
where we changed variables to $\tilde{r}_s = \tilde{r}'-\tilde{r}''$, $R = \tilde{r}'/2+\tilde{r}''/2$ and expanded about small $\tilde{r}_s$. We note that the $\tilde{r}_s$ integral with the cutoff $\tilde{r}_s>\alpha$ is rotationally symmetric and thus the cross terms are zero. Expanding the square, integrating by parts, and using the fact that $\Delta_R F(R) = -2\pi \delta^{(2)}(R)$ gives
\begin{align}
R^{\rho+}_{a}(\tilde{r})^{(2)}_{g_1} =& -m^2 a^2K_{\rho+}^2 e^{-a^2 K_{\rho+}F(\tilde{r})/2} \int \frac{\td^2 \tilde{r}_s \td^2 R}{(2\pi \alpha)^4 v^2} e^{-2(m^2 K_{\rho+} + K_{\sigma+})F(\tilde{r}_s)} \tilde{r}_s^2\left[F(\tilde{r}-R)-F(R)\right]\Delta_R\left[F(\tilde{r}-R)-F(R)\right]\\
=& \frac{m^2 a^2K_{\rho+}^2}{\pi v^2}F(\tilde{r}) e^{-a^2 K_{\rho+}F(\tilde{r})/2} \int \frac{\td^2 \tilde{r}_s}{4\pi^2 \alpha^4} \tilde{r}_s^2 e^{-2(m^2 K_{\rho+} + K_{\sigma+})F(\tilde{r}_s)}\\
=& \frac{a^2}{2} \frac{1}{\pi^2 v^2}m^2 K_{\rho+}^2F(\tilde{r}) e^{-a^2 K_{\rho+}F(\tilde{r})/2} \int_{\tilde{r}_s>\alpha} \frac{\td \tilde{r}_s}{\alpha}\ \left(\frac{\tilde{r}_s}{\alpha}\right)^{3-2(m^2K_{\rho+}+K_{\sigma+})}.
\end{align}
Computing the second-order correlator for $g_2$ will be essentially identical. Note that there are no cross terms involving $g_1g_2$ because the spin parts of their perturbations must have canceling exponents and each spin piece is distinct. In the end, we will see that
\begin{align}
R^{\rho+}_a(\tilde{r}) =& e^{-a^2K_{\rho+}F(\tilde{r})/2}\left\{1 + \frac{a^2}{2}m^2 K_{\rho+}^2 F(\tilde{r}) \int_{\tilde{r}_s>\alpha} \frac{\td \tilde{r}_s}{\alpha} \ \left[y_1^2(\alpha)\left(\frac{\tilde{r}_s}{\alpha}\right)^{3-2(m^2K_{\rho+}+K_{\sigma+})} + y_2^2(\alpha)\left(\frac{\tilde{r}_s}{\alpha}\right)^{3-2(m^2K_{\rho+}+K_{\sigma-})}\right]\right\},
\end{align}
where $y_i = g_i/(\sqrt{2}\pi v)$. Demanding that the effective value of $K_{\rho+}$ is invariant under changing $\alpha$ to $\alpha e^{\td l}$ gives us the RG equations:
\begin{align}
\frac{\td K_{\rho+}}{\td l} =& -(y_1^2 + y_2^2)m^2 K_{\rho+}^2,\\
\frac{\td y_1}{\td l} =& (2 - m^2K_{\rho+} - K_{\sigma+})y_1,\\
\frac{\td y_2}{\td l} =& (2 - m^2K_{\rho+} - K_{\sigma-})y_2.
\end{align}
These latter two RG equations follow directly from the scaling dimensions of the $g_i$ terms. It is only the first that is nontrivial. 

We will next compute $R_a^{\sigma+}(\tilde{r})$. We note that the $\theta_{\sigma+}$ mode can only couple to the $g_1$ term, so
\begin{align}
R_a^{\sigma+}(\tilde{r}) =& e^{-a^2K_{\sigma+}F(\tilde{r})/2} + \frac{1}{2}g_1^2 R_a^{\sigma+}(\tilde{r})^{(2)}_{g_1},\\
R_a^{\sigma+}(\tilde{r})^{(2)}_{g_1} =& 2e^{-a^2K_{\sigma+}F(\tilde{r})/2}\sum_\sigma \int \frac{\td^2 \tilde{r}'\td^2 \tilde{r}''}{(2\pi \alpha)^4 v^2}e^{-2(m^2K_{\rho+}+K_{\sigma+})F(\tilde{r}'-\tilde{r}'')}\left[e^{a\sigma K_{\sigma+}[F(\tilde{r}-\tilde{r}')-F(\tilde{r}-\tilde{r}'')+F(\tilde{r}'') - F(\tilde{r}')]}-1\right]\\
=& \frac{a^2}{2}\frac{1}{\pi^2 v^2} K_{\sigma+}^2F(\tilde{r}) e^{-a^2K_{\sigma+}F(\tilde{r})/2}\int_{\tilde{r}_s>\alpha} \frac{\td \tilde{r}_s}{\alpha} \ \left(\frac{\tilde{r}_s}{\alpha}\right)^{3 - 2(m^2 K_{\rho+}+K_{\sigma+})},
\end{align}
where we have again used the same tricks in our computation of these correlators and expansions. Finding the effective $K_{\sigma+}$ and the RG equations is again straightforward. We see that we will have
\begin{equation}
\frac{\td K_{\sigma+}}{\td l} = -y_1^2 K_{\sigma+}^2,
\end{equation}
and the identical RG equation we have already found for $y_1$. Since $g_3=0$, the theory is symmetric under the interchange $\theta_{\sigma+}\leftrightarrow \theta_{\sigma-}, g_1\leftrightarrow g_2$. Thus, the last RG equation is clearly
\begin{equation}
\frac{\td K_{\sigma-}}{\td l} = -y_2^2 K_{\sigma-}^2.
\end{equation}

We can now linearize our RG equations about the fixed point. We know that if we define $x_{\sigma\pm} = 1 - K_{\sigma\pm}$, then $x_{\sigma\pm}$ will be zero in the $\mathrm{SU}(2)$ invariant C2S2 theory, and should be expected to be small at the critical point. If we also define $x_{\rho+} = 1-m^2K_{\rho+}$, then our four RG equations will become
\begin{align}
\frac{\td x_{\rho+}}{\td l} =& y_1^2+y_2^2,\\
\frac{\td x_{\sigma+}}{\td l} =& y_1^2,\\
\frac{\td x_{\sigma-}}{\td l} =& y_2^2,\\
\frac{\td y_1}{\td l} =& (x_{\rho+} + x_{\sigma+})y_1,\\
\frac{\td y_2}{\td l} =& (x_{\rho+} + x_{\sigma-})y_2.
\end{align}
The interpretation of these RG equations is straightforward. The presence of any of the two terms will attempt to pin $\theta_{\rho+}$, which will increase $x_{\rho+}$ (decrease $K_{\rho+})$. Due to the symmetry of the problem, they must all contribute the same way. The first term will also attempt to pin $\theta_{\sigma+}$, while the last will attempt to pin $\theta_{\sigma-}$.

Let $f_i = \ln(y_i)$, then we see that $\td(f_1-f_2)/\td l = x_{\sigma+} - x_{\sigma-}$ and $\td(x_{\sigma+} - x_{\sigma-})/\td l = e^{2f_1}-e^{2f_2}$. From this, we see that $\td^2(f_1-f_2)/\td l^2 = e^{2f_1}-e^{2f_2}$ and thus that $\td^2 f_1/\td l^2 - e^{2f_1} = g(l) = \td^2 f_2/\td l^2 -e^{2f_2}$, where $g(l)$ is some function of $l$. We know from the existence and uniqueness theorem that the equation $\td^2 f_1/\td l^2 - e^{2f_1} = g(l)$ has a unique solution for a given initial condition $f_1(l=0) = f_{1,0}$ and $\td f_1(l=0)/\td l = f'_{1,0}$. Now, at the beginning of our flow, we have that $y_1(l=0) = y_u = y_2(l=0)$, since both come from the umklapp operator. Further, we start from the $
\mathrm{SU}(2)$ invariant fixed point in the C2S2 theory, where $x_{\sigma+} = 0 = x_{\sigma-}$. Thus we conclude that $\td f_1(l=0)/\td l = x_{\rho+}(l=0) = \td f_2(l=0)/\td l$. Then, since $f_1, f_2$ satisfy the same second-order differential equation and have the same two initial conditions, we can conclude that $y_1=y_2$ by the uniqueness theorem. 

Our equations then reduce further to
\begin{align}
\frac{\td x_{\rho+}}{\td l} =& 2y_u^2,\\
\frac{\td x_{\sigma+}}{\td l} =& y_u^2 = \frac{\td x_{\sigma-}}{\td l},\\
\frac{\td y_u}{\td l} =& (x_{\rho+} + x_{\sigma+})y_u,
\end{align}
where we can conclude that $x_{\sigma+}(l) = x_{\sigma-}(l)$ and both are zero initially. The first two equations will also reveal that $x_{\rho+}(l) = 2x_{\sigma+}(l) + x_{\rho+,0}$. But then the above flow can immediately be seen to be the traditional KT-like transition. It is controlled by only two variables $y_u$ and $x_{\rho+}$, with all other values fixed by $\mathrm{SU}(2)$ invariance and initial conditions.

Consider the flow in the disordered regime. We can take advantage of the fact that $A^2 = 3y^2 - (x_{\rho+}+x_{\sigma+})^2$ is a constant of the flow; the disordered regime has this constant being greater than zero. We can then solve the above equations to find that
\begin{align}
x_{\rho+}(l) =& \frac{2}{3}A\tan\left[Al + \tan^{-1}\left(\frac{x_{\rho+,0}}{A}\right)\right] + \frac{1}{3}x_{\rho+,0},\\
x_{\sigma+}(l) =& \frac{1}{3}A\tan\left[Al + \tan^{-1}\left(\frac{x_{\rho+,0}}{A}\right)\right] - \frac{1}{3}x_{\rho+,0},\\
y_u(l) =& \frac{1}{\sqrt{3}}A\sec\left[Al + \tan^{-1}\left(\frac{x_{\rho+,0}}{A}\right)\right].
\end{align}
For small $A$ and small $l$, these equations reduce to those of the separatrix, as discussed in the main text. 

We can now understand the scaling of the gaps as we approach the critical point. For ease, we will choose $x_{\rho+,0}=0$ so $A = \sqrt{3}|y_0|$ and we will tune $A$ towards zero. On dimensional grounds \cite{giamarchi_quantum_2003}, we can expect that $\Delta_{c/s} \sim e^{l^*_{c/s}}$ where $x_{\rho+}(l^*_c)\sim 1$ and $x_{\sigma\pm}(l^*_s)\sim 1$. Then we see that $3/(2A) = \tan(Al^*_c)$ and $3/A = \tan(Al^*_s)$. For small $A$, we can expand this to see that $l^*_c \sim \pi/(2A) - 2/3$ and $l^*_s \sim \pi/(2A) - 1/3$ and thus $l^*_c\sim l^*_s$. In particular, $A\sim \sqrt{2-m^2K_{\rho+}-K_{\sigma+}}$ if the transition is approached from a different direction, giving the scaling of $\Delta_c$ and $\Delta_s$ quoted in the main text.

Lastly, we address the scaling of the correlator $\langle \sc{O}_{\mathrm{CDW}}(\tilde{r})\sc{O}_{\mathrm{CDW}}(0)\rangle$. If we make the same definition as in the above, $R_a^{\rho+}(\tilde{r},\alpha) = I_a(\alpha,\alpha')R_a^{\rho+}(\tilde{r},\alpha')$ then we can use our derivation of the correlator above to see that
\begin{align}
I_a(\alpha,\alpha e^{\td l}) =& 1 + \frac{a^2}{2}m^2 K_{\rho+}^2(\alpha)[y_1^2(\alpha) + y_2^2(\alpha)] F(\tilde{r}).
\end{align}
By the identical logic as in the even $m$ case, we see that therefore
\begin{align}
R_a^{\rho+}(\tilde{r},\alpha) =& \exp\left[-\frac{a^2}{2}\int^{\ln(\tilde{r}/\alpha)}_0 \td l \ K_{\rho+}(l)\right].
\end{align}
Close to the critical point, we will then have that
\begin{align}
R_a^{\rho+}(\tilde{r},\alpha) =& \left(\frac{\alpha}{\tilde{r}}\right)^{a^2/2m^2}\exp\left[\frac{a^2}{2m^2}\int^{\ln(\tilde{r}/\alpha)}_0 \td l \ \left(\frac{2}{3}\frac{x_{\rho+,0}}{1-x_{\rho+,0}l} + \frac{1}{3}x_{\rho+,0}\right)\right].
\end{align}
If we use the definition of $\xi_c = \alpha e^{l^*_c}$ with $x_{\rho+}(l^*_c) = 1$, then we can evaluate this integral to find that
\begin{equation}
R^{\rho+}_a(\xi,\alpha) = \left(\frac{\alpha}{\xi}\right)^{a^2/2m^2}x_{\rho+,0}^{-a^2/3m^2},
\end{equation}
where this power is due to the factor of $2/3$ in front of the diverging piece of $x_{\rho+}(l)$. We finally conclude that we should expect
\begin{equation}
\langle \sc{O}_{\mathrm{CDW}}\rangle \sim \exp\left(-\frac{4C/m^2}{\sqrt{2-m^2K_{\rho+}-K_{\sigma+}}}\right)\left(2-m^2K_{\rho+}-K_{\sigma+}\right)^{-8/3m^2}.
\end{equation}

\twocolumngrid

\bibliographystyle{apsrev4-1_custom}
\bibliography{triangle_strip.bib}

\begin{thebibliography}{23}%
\makeatletter
\providecommand \@ifxundefined [1]{%
 \@ifx{#1\undefined}
}%
\providecommand \@ifnum [1]{%
 \ifnum #1\expandafter \@firstoftwo
 \else \expandafter \@secondoftwo
 \fi
}%
\providecommand \@ifx [1]{%
 \ifx #1\expandafter \@firstoftwo
 \else \expandafter \@secondoftwo
 \fi
}%
\providecommand \natexlab [1]{#1}%
\providecommand \enquote  [1]{``#1''}%
\providecommand \bibnamefont  [1]{#1}%
\providecommand \bibfnamefont [1]{#1}%
\providecommand \citenamefont [1]{#1}%
\providecommand \href@noop [0]{\@secondoftwo}%
\providecommand \href [0]{\begingroup \@sanitize@url \@href}%
\providecommand \@href[1]{\@@startlink{#1}\@@href}%
\providecommand \@@href[1]{\endgroup#1\@@endlink}%
\providecommand \@sanitize@url [0]{\catcode `\\12\catcode `\$12\catcode
  `\&12\catcode `\#12\catcode `\^12\catcode `\_12\catcode `\%12\relax}%
\providecommand \@@startlink[1]{}%
\providecommand \@@endlink[0]{}%
\providecommand \url  [0]{\begingroup\@sanitize@url \@url }%
\providecommand \@url [1]{\endgroup\@href {#1}{\urlprefix }}%
\providecommand \urlprefix  [0]{URL }%
\providecommand \Eprint [0]{\href }%
\providecommand \doibase [0]{http://dx.doi.org/}%
\providecommand \selectlanguage [0]{\@gobble}%
\providecommand \bibinfo  [0]{\@secondoftwo}%
\providecommand \bibfield  [0]{\@secondoftwo}%
\providecommand \translation [1]{[#1]}%
\providecommand \BibitemOpen [0]{}%
\providecommand \bibitemStop [0]{}%
\providecommand \bibitemNoStop [0]{.\EOS\space}%
\providecommand \EOS [0]{\spacefactor3000\relax}%
\providecommand \BibitemShut  [1]{\csname bibitem#1\endcsname}%
\let\auto@bib@innerbib\@empty
%</preamble>
\bibitem [{\citenamefont
  {Senthil}(2008{\natexlab{a}})}]{senthil_critical_2008}%
  \BibitemOpen
  \bibfield  {author} {\bibinfo {author} {\bibfnamefont {T.}~\bibnamefont
  {Senthil}},\ }\bibfield  {title} {\enquote {\bibinfo {title} {Critical
  {Fermi} surfaces and non-{Fermi} liquid metals},}\ }\href {\doibase
  10.1103/PhysRevB.78.035103} {\bibfield  {journal} {\bibinfo  {journal}
  {Physical Review B}\ }\textbf {\bibinfo {volume} {78}},\ \bibinfo {pages}
  {035103} (\bibinfo {year} {2008}{\natexlab{a}})}\BibitemShut {NoStop}%
\bibitem [{\citenamefont {Li}\ \emph {et~al.}(2021{\natexlab{a}})\citenamefont
  {Li}, \citenamefont {Jiang}, \citenamefont {Li}, \citenamefont {Zhang},
  \citenamefont {Kang}, \citenamefont {Zhu}, \citenamefont {Watanabe},
  \citenamefont {Taniguchi}, \citenamefont {Chowdhury}, \citenamefont {Fu},
  \citenamefont {Shan},\ and\ \citenamefont {Mak}}]{li_continuous_2021}%
  \BibitemOpen
  \bibfield  {author} {\bibinfo {author} {\bibfnamefont {T.}~\bibnamefont
  {Li}}, \bibinfo {author} {\bibfnamefont {S.}~\bibnamefont {Jiang}}, \bibinfo
  {author} {\bibfnamefont {L.}~\bibnamefont {Li}}, \bibinfo {author}
  {\bibfnamefont {Y.}~\bibnamefont {Zhang}}, \bibinfo {author} {\bibfnamefont
  {K.}~\bibnamefont {Kang}}, \bibinfo {author} {\bibfnamefont {J.}~\bibnamefont
  {Zhu}}, \bibinfo {author} {\bibfnamefont {K.}~\bibnamefont {Watanabe}},
  \bibinfo {author} {\bibfnamefont {T.}~\bibnamefont {Taniguchi}}, \bibinfo
  {author} {\bibfnamefont {D.}~\bibnamefont {Chowdhury}}, \bibinfo {author}
  {\bibfnamefont {L.}~\bibnamefont {Fu}}, \bibinfo {author} {\bibfnamefont
  {J.}~\bibnamefont {Shan}}, \ and\ \bibinfo {author} {\bibfnamefont {K.~F.}\
  \bibnamefont {Mak}},\ }\bibfield  {title} {\enquote {\bibinfo {title}
  {Continuous {Mott} transition in semiconductor moir\'{e} superlattices},}\
  }\href {\doibase 10.1038/s41586-021-03853-0} {\bibfield  {journal} {\bibinfo
  {journal} {Nature}\ }\textbf {\bibinfo {volume} {597}},\ \bibinfo {pages}
  {350} (\bibinfo {year} {2021}{\natexlab{a}})}\BibitemShut {NoStop}%
\bibitem [{\citenamefont {Ghiotto}\ \emph {et~al.}(2021)\citenamefont
  {Ghiotto}, \citenamefont {Shih}, \citenamefont {Pereira}, \citenamefont
  {Rhodes}, \citenamefont {Kim}, \citenamefont {Zang}, \citenamefont {Millis},
  \citenamefont {Watanabe}, \citenamefont {Taniguchi}, \citenamefont {Hone},
  \citenamefont {Wang}, \citenamefont {Dean},\ and\ \citenamefont
  {Pasupathy}}]{ghiotto_quantum_2021}%
  \BibitemOpen
  \bibfield  {author} {\bibinfo {author} {\bibfnamefont {A.}~\bibnamefont
  {Ghiotto}}, \bibinfo {author} {\bibfnamefont {E.-M.}\ \bibnamefont {Shih}},
  \bibinfo {author} {\bibfnamefont {G.~S. S.~G.}\ \bibnamefont {Pereira}},
  \bibinfo {author} {\bibfnamefont {D.~A.}\ \bibnamefont {Rhodes}}, \bibinfo
  {author} {\bibfnamefont {B.}~\bibnamefont {Kim}}, \bibinfo {author}
  {\bibfnamefont {J.}~\bibnamefont {Zang}}, \bibinfo {author} {\bibfnamefont
  {A.~J.}\ \bibnamefont {Millis}}, \bibinfo {author} {\bibfnamefont
  {K.}~\bibnamefont {Watanabe}}, \bibinfo {author} {\bibfnamefont
  {T.}~\bibnamefont {Taniguchi}}, \bibinfo {author} {\bibfnamefont {J.~C.}\
  \bibnamefont {Hone}}, \bibinfo {author} {\bibfnamefont {L.}~\bibnamefont
  {Wang}}, \bibinfo {author} {\bibfnamefont {C.~R.}\ \bibnamefont {Dean}}, \
  and\ \bibinfo {author} {\bibfnamefont {A.~N.}\ \bibnamefont {Pasupathy}},\
  }\bibfield  {title} {\enquote {\bibinfo {title} {Quantum criticality in
  twisted transition metal dichalcogenides},}\ }\href {\doibase
  10.1038/s41586-021-03815-6} {\bibfield  {journal} {\bibinfo  {journal}
  {Nature}\ }\textbf {\bibinfo {volume} {597}},\ \bibinfo {pages} {345}
  (\bibinfo {year} {2021})}\BibitemShut {NoStop}%
\bibitem [{\citenamefont {Senthil}(2008{\natexlab{b}})}]{senthil_theory_2008}%
  \BibitemOpen
  \bibfield  {author} {\bibinfo {author} {\bibfnamefont {T.}~\bibnamefont
  {Senthil}},\ }\bibfield  {title} {\enquote {\bibinfo {title} {Theory of a
  continuous {Mott} transition in two dimensions},}\ }\href {\doibase
  10.1103/PhysRevB.78.045109} {\bibfield  {journal} {\bibinfo  {journal}
  {Physical Review B}\ }\textbf {\bibinfo {volume} {78}},\ \bibinfo {pages}
  {045109} (\bibinfo {year} {2008}{\natexlab{b}})}\BibitemShut {NoStop}%
\bibitem [{\citenamefont {Kim}\ \emph {et~al.}(2022)\citenamefont {Kim},
  \citenamefont {Senthil},\ and\ \citenamefont
  {Chowdhury}}]{kim_continuous_2022}%
  \BibitemOpen
  \bibfield  {author} {\bibinfo {author} {\bibfnamefont {S.}~\bibnamefont
  {Kim}}, \bibinfo {author} {\bibfnamefont {T.}~\bibnamefont {Senthil}}, \ and\
  \bibinfo {author} {\bibfnamefont {D.}~\bibnamefont {Chowdhury}},\ }\bibfield
  {title} {\enquote {\bibinfo {title} {Continuous {Mott} transition in
  moir{\textbackslash}'e semiconductors: role of long-wavelength
  inhomogeneities},}\ }\href {http://arxiv.org/abs/2204.10865} {\bibfield
  {journal} {\bibinfo  {journal} {arXiv:2204.10865 [cond-mat]}\ } (\bibinfo
  {year} {2022})},\ \bibinfo {note} {arXiv: 2204.10865}\BibitemShut {NoStop}%
\bibitem [{\citenamefont {Lee}\ and\ \citenamefont
  {Senthil}()}]{unpublishedSenthil}%
  \BibitemOpen
  \bibfield  {author} {\bibinfo {author} {\bibfnamefont {P.}~\bibnamefont
  {Lee}}\ and\ \bibinfo {author} {\bibfnamefont {T.}~\bibnamefont {Senthil}},\
  }\href@noop {} {}\bibinfo {note} {(unpublished)}\BibitemShut {NoStop}%
\bibitem [{\citenamefont {Tang}\ \emph {et~al.}(2022)\citenamefont {Tang},
  \citenamefont {Gu}, \citenamefont {Liu}, \citenamefont {Watanabe},
  \citenamefont {Taniguchi}, \citenamefont {Hone}, \citenamefont {Mak},\ and\
  \citenamefont {Shan}}]{tang_dielectric_2022}%
  \BibitemOpen
  \bibfield  {author} {\bibinfo {author} {\bibfnamefont {Y.}~\bibnamefont
  {Tang}}, \bibinfo {author} {\bibfnamefont {J.}~\bibnamefont {Gu}}, \bibinfo
  {author} {\bibfnamefont {S.}~\bibnamefont {Liu}}, \bibinfo {author}
  {\bibfnamefont {K.}~\bibnamefont {Watanabe}}, \bibinfo {author}
  {\bibfnamefont {T.}~\bibnamefont {Taniguchi}}, \bibinfo {author}
  {\bibfnamefont {J.~C.}\ \bibnamefont {Hone}}, \bibinfo {author}
  {\bibfnamefont {K.~F.}\ \bibnamefont {Mak}}, \ and\ \bibinfo {author}
  {\bibfnamefont {J.}~\bibnamefont {Shan}},\ }\bibfield  {title} {\enquote
  {\bibinfo {title} {Dielectric catastrophe at the {Wigner}-{Mott} transition
  in a moiré superlattice},}\ }\href {\doibase 10.1038/s41467-022-32037-1}
  {\bibfield  {journal} {\bibinfo  {journal} {Nature Communications}\ }\textbf
  {\bibinfo {volume} {13}},\ \bibinfo {pages} {4271} (\bibinfo {year}
  {2022})}\BibitemShut {NoStop}%
\bibitem [{\citenamefont {Mak}\ and\ \citenamefont {Shan}()}]{MakShan}%
  \BibitemOpen
  \bibfield  {author} {\bibinfo {author} {\bibfnamefont {K.~F.}\ \bibnamefont
  {Mak}}\ and\ \bibinfo {author} {\bibfnamefont {J.}~\bibnamefont {Shan}},\
  }\href@noop {} {}\bibinfo {howpublished} {private communication}\BibitemShut
  {NoStop}%
\bibitem [{\citenamefont {Xu}\ \emph {et~al.}(2022)\citenamefont {Xu},
  \citenamefont {Wu}, \citenamefont {Ye}, \citenamefont {Luo}, \citenamefont
  {Jian},\ and\ \citenamefont {Xu}}]{xu_interaction-driven_2022}%
  \BibitemOpen
  \bibfield  {author} {\bibinfo {author} {\bibfnamefont {Y.}~\bibnamefont
  {Xu}}, \bibinfo {author} {\bibfnamefont {X.-C.}\ \bibnamefont {Wu}}, \bibinfo
  {author} {\bibfnamefont {M.}~\bibnamefont {Ye}}, \bibinfo {author}
  {\bibfnamefont {Z.-X.}\ \bibnamefont {Luo}}, \bibinfo {author} {\bibfnamefont
  {C.-M.}\ \bibnamefont {Jian}}, \ and\ \bibinfo {author} {\bibfnamefont
  {C.}~\bibnamefont {Xu}},\ }\bibfield  {title} {\enquote {\bibinfo {title}
  {Interaction-{Driven} {Metal}-{Insulator} {Transition} with {Charge}
  {Fractionalization}},}\ }\href {\doibase 10.1103/PhysRevX.12.021067}
  {\bibfield  {journal} {\bibinfo  {journal} {Physical Review X}\ }\textbf
  {\bibinfo {volume} {12}},\ \bibinfo {pages} {021067} (\bibinfo {year}
  {2022})}\BibitemShut {NoStop}%
\bibitem [{\citenamefont {Musser}\ \emph {et~al.}(2022)\citenamefont {Musser},
  \citenamefont {Senthil},\ and\ \citenamefont
  {Chowdhury}}]{musser_theory_2022}%
  \BibitemOpen
  \bibfield  {author} {\bibinfo {author} {\bibfnamefont {S.}~\bibnamefont
  {Musser}}, \bibinfo {author} {\bibfnamefont {T.}~\bibnamefont {Senthil}}, \
  and\ \bibinfo {author} {\bibfnamefont {D.}~\bibnamefont {Chowdhury}},\
  }\bibfield  {title} {\enquote {\bibinfo {title} {Theory of a continuous
  bandwidth-tuned {Wigner}-{Mott} transition},}\ }\href {\doibase
  10.1103/PhysRevB.106.155145} {\bibfield  {journal} {\bibinfo  {journal}
  {Physical Review B}\ }\textbf {\bibinfo {volume} {106}},\ \bibinfo {pages}
  {155145} (\bibinfo {year} {2022})}\BibitemShut {NoStop}%
\bibitem [{\citenamefont {Schulz}(1994)}]{schulz_metal-insulator_1994}%
  \BibitemOpen
  \bibfield  {author} {\bibinfo {author} {\bibfnamefont {H.~J.}\ \bibnamefont
  {Schulz}},\ }\bibfield  {title} {\enquote {\bibinfo {title} {The
  metal-insulator transition in one dimension},}\ }\href
  {http://arxiv.org/abs/cond-mat/9412036} {\bibfield  {journal} {\bibinfo
  {journal} {arXiv:cond-mat/9412036}\ } (\bibinfo {year} {1994})},\ \bibinfo
  {note} {arXiv: cond-mat/9412036}\BibitemShut {NoStop}%
\bibitem [{\citenamefont {Giamarchi}(2003)}]{giamarchi_quantum_2003}%
  \BibitemOpen
  \bibfield  {author} {\bibinfo {author} {\bibfnamefont {T.}~\bibnamefont
  {Giamarchi}},\ }\href {\doibase 10.1093/acprof:oso/9780198525004.001.0001}
  {\emph {\bibinfo {title} {Quantum {Physics} in {One} {Dimension}}}}\
  (\bibinfo  {publisher} {Oxford University Press},\ \bibinfo {year}
  {2003})\BibitemShut {NoStop}%
\bibitem [{\citenamefont {Mishmash}\ \emph {et~al.}(2015)\citenamefont
  {Mishmash}, \citenamefont {Gonz\'{a}lez}, \citenamefont {Melko},
  \citenamefont {Motrunich},\ and\ \citenamefont
  {Fisher}}]{mishmash_continuous_2015}%
  \BibitemOpen
  \bibfield  {author} {\bibinfo {author} {\bibfnamefont {R.~V.}\ \bibnamefont
  {Mishmash}}, \bibinfo {author} {\bibfnamefont {I.}~\bibnamefont
  {Gonz\'{a}lez}}, \bibinfo {author} {\bibfnamefont {R.~G.}\ \bibnamefont
  {Melko}}, \bibinfo {author} {\bibfnamefont {O.~I.}\ \bibnamefont
  {Motrunich}}, \ and\ \bibinfo {author} {\bibfnamefont {M.~P.~A.}\
  \bibnamefont {Fisher}},\ }\bibfield  {title} {\enquote {\bibinfo {title}
  {Continuous {Mott} transition between a metal and a quantum spin liquid},}\
  }\href {\doibase 10.1103/PhysRevB.91.235140} {\bibfield  {journal} {\bibinfo
  {journal} {Physical Review B}\ }\textbf {\bibinfo {volume} {91}},\ \bibinfo
  {pages} {235140} (\bibinfo {year} {2015})}\BibitemShut {NoStop}%
\bibitem [{\citenamefont {Balents}\ and\ \citenamefont
  {Fisher}(1996)}]{balents_weak-coupling_1996}%
  \BibitemOpen
  \bibfield  {author} {\bibinfo {author} {\bibfnamefont {L.}~\bibnamefont
  {Balents}}\ and\ \bibinfo {author} {\bibfnamefont {M.~P.~A.}\ \bibnamefont
  {Fisher}},\ }\bibfield  {title} {\enquote {\bibinfo {title} {Weak-coupling
  phase diagram of the two-chain {Hubbard} model},}\ }\href {\doibase
  10.1103/PhysRevB.53.12133} {\bibfield  {journal} {\bibinfo  {journal}
  {Physical Review B}\ }\textbf {\bibinfo {volume} {53}},\ \bibinfo {pages}
  {12133} (\bibinfo {year} {1996})}\BibitemShut {NoStop}%
\bibitem [{\citenamefont {Sheng}\ \emph {et~al.}(2009)\citenamefont {Sheng},
  \citenamefont {Motrunich},\ and\ \citenamefont {Fisher}}]{sheng_spin_2009}%
  \BibitemOpen
  \bibfield  {author} {\bibinfo {author} {\bibfnamefont {D.~N.}\ \bibnamefont
  {Sheng}}, \bibinfo {author} {\bibfnamefont {O.~I.}\ \bibnamefont
  {Motrunich}}, \ and\ \bibinfo {author} {\bibfnamefont {M.~P.~A.}\
  \bibnamefont {Fisher}},\ }\bibfield  {title} {\enquote {\bibinfo {title}
  {Spin {Bose}-metal phase in a spin- 1 2 model with ring exchange on a two-leg
  triangular strip},}\ }\href {\doibase 10.1103/PhysRevB.79.205112} {\bibfield
  {journal} {\bibinfo  {journal} {Physical Review B}\ }\textbf {\bibinfo
  {volume} {79}},\ \bibinfo {pages} {205112} (\bibinfo {year}
  {2009})}\BibitemShut {NoStop}%
\bibitem [{\citenamefont {Xu}\ \emph {et~al.}(2020)\citenamefont {Xu},
  \citenamefont {Liu}, \citenamefont {Rhodes}, \citenamefont {Watanabe},
  \citenamefont {Taniguchi}, \citenamefont {Hone}, \citenamefont {Elser},
  \citenamefont {Mak},\ and\ \citenamefont {Shan}}]{xu_correlated_2020}%
  \BibitemOpen
  \bibfield  {author} {\bibinfo {author} {\bibfnamefont {Y.}~\bibnamefont
  {Xu}}, \bibinfo {author} {\bibfnamefont {S.}~\bibnamefont {Liu}}, \bibinfo
  {author} {\bibfnamefont {D.~A.}\ \bibnamefont {Rhodes}}, \bibinfo {author}
  {\bibfnamefont {K.}~\bibnamefont {Watanabe}}, \bibinfo {author}
  {\bibfnamefont {T.}~\bibnamefont {Taniguchi}}, \bibinfo {author}
  {\bibfnamefont {J.}~\bibnamefont {Hone}}, \bibinfo {author} {\bibfnamefont
  {V.}~\bibnamefont {Elser}}, \bibinfo {author} {\bibfnamefont {K.~F.}\
  \bibnamefont {Mak}}, \ and\ \bibinfo {author} {\bibfnamefont
  {J.}~\bibnamefont {Shan}},\ }\bibfield  {title} {\enquote {\bibinfo {title}
  {Correlated insulating states at fractional fillings of moir\'{e}
  superlattices},}\ }\href {\doibase 10.1038/s41586-020-2868-6} {\bibfield
  {journal} {\bibinfo  {journal} {Nature}\ }\textbf {\bibinfo {volume} {587}},\
  \bibinfo {pages} {214} (\bibinfo {year} {2020})}\BibitemShut {NoStop}%
\bibitem [{\citenamefont {Li}\ \emph {et~al.}(2021{\natexlab{b}})\citenamefont
  {Li}, \citenamefont {Li}, \citenamefont {Regan}, \citenamefont {Wang},
  \citenamefont {Zhao}, \citenamefont {Kahn}, \citenamefont {Yumigeta},
  \citenamefont {Blei}, \citenamefont {Taniguchi}, \citenamefont {Watanabe},
  \citenamefont {Tongay}, \citenamefont {Zettl}, \citenamefont {Crommie},\ and\
  \citenamefont {Wang}}]{li_imaging_2021}%
  \BibitemOpen
  \bibfield  {author} {\bibinfo {author} {\bibfnamefont {H.}~\bibnamefont
  {Li}}, \bibinfo {author} {\bibfnamefont {S.}~\bibnamefont {Li}}, \bibinfo
  {author} {\bibfnamefont {E.~C.}\ \bibnamefont {Regan}}, \bibinfo {author}
  {\bibfnamefont {D.}~\bibnamefont {Wang}}, \bibinfo {author} {\bibfnamefont
  {W.}~\bibnamefont {Zhao}}, \bibinfo {author} {\bibfnamefont {S.}~\bibnamefont
  {Kahn}}, \bibinfo {author} {\bibfnamefont {K.}~\bibnamefont {Yumigeta}},
  \bibinfo {author} {\bibfnamefont {M.}~\bibnamefont {Blei}}, \bibinfo {author}
  {\bibfnamefont {T.}~\bibnamefont {Taniguchi}}, \bibinfo {author}
  {\bibfnamefont {K.}~\bibnamefont {Watanabe}}, \bibinfo {author}
  {\bibfnamefont {S.}~\bibnamefont {Tongay}}, \bibinfo {author} {\bibfnamefont
  {A.}~\bibnamefont {Zettl}}, \bibinfo {author} {\bibfnamefont {M.~F.}\
  \bibnamefont {Crommie}}, \ and\ \bibinfo {author} {\bibfnamefont
  {F.}~\bibnamefont {Wang}},\ }\bibfield  {title} {\enquote {\bibinfo {title}
  {Imaging two-dimensional generalized {Wigner} crystals},}\ }\href {\doibase
  10.1038/s41586-021-03874-9} {\bibfield  {journal} {\bibinfo  {journal}
  {Nature}\ }\textbf {\bibinfo {volume} {597}},\ \bibinfo {pages} {650}
  (\bibinfo {year} {2021}{\natexlab{b}})}\BibitemShut {NoStop}%
\bibitem [{\citenamefont {Chitra}\ \emph {et~al.}(1995)\citenamefont {Chitra},
  \citenamefont {Pati}, \citenamefont {Krishnamurthy}, \citenamefont {Sen},\
  and\ \citenamefont {Ramasesha}}]{chitra_density-matrix_1995}%
  \BibitemOpen
  \bibfield  {author} {\bibinfo {author} {\bibfnamefont {R.}~\bibnamefont
  {Chitra}}, \bibinfo {author} {\bibfnamefont {S.}~\bibnamefont {Pati}},
  \bibinfo {author} {\bibfnamefont {H.~R.}\ \bibnamefont {Krishnamurthy}},
  \bibinfo {author} {\bibfnamefont {D.}~\bibnamefont {Sen}}, \ and\ \bibinfo
  {author} {\bibfnamefont {S.}~\bibnamefont {Ramasesha}},\ }\bibfield  {title}
  {\enquote {\bibinfo {title} {Density-matrix renormalization-group studies of
  the spin-1/2 {Heisenberg} system with dimerization and frustration},}\ }\href
  {\doibase 10.1103/PhysRevB.52.6581} {\bibfield  {journal} {\bibinfo
  {journal} {Physical Review B}\ }\textbf {\bibinfo {volume} {52}},\ \bibinfo
  {pages} {6581} (\bibinfo {year} {1995})}\BibitemShut {NoStop}%
\bibitem [{\citenamefont {White}\ and\ \citenamefont
  {Affleck}(1996)}]{white_dimerization_1996}%
  \BibitemOpen
  \bibfield  {author} {\bibinfo {author} {\bibfnamefont {S.~R.}\ \bibnamefont
  {White}}\ and\ \bibinfo {author} {\bibfnamefont {I.}~\bibnamefont
  {Affleck}},\ }\bibfield  {title} {\enquote {\bibinfo {title} {Dimerization
  and incommensurate spiral spin correlations in the zigzag spin chain:
  {Analogies} to the {Kondo} lattice},}\ }\href {\doibase
  10.1103/PhysRevB.54.9862} {\bibfield  {journal} {\bibinfo  {journal}
  {Physical Review B}\ }\textbf {\bibinfo {volume} {54}},\ \bibinfo {pages}
  {9862} (\bibinfo {year} {1996})}\BibitemShut {NoStop}%
\bibitem [{\citenamefont {Hida}(1992)}]{hida_crossover_1992}%
  \BibitemOpen
  \bibfield  {author} {\bibinfo {author} {\bibfnamefont {K.}~\bibnamefont
  {Hida}},\ }\bibfield  {title} {\enquote {\bibinfo {title} {Crossover between
  the {Haldane}-gap phase and the dimer phase in the spin-1/2 alternating
  {Heisenberg} chain},}\ }\href {\doibase 10.1103/PhysRevB.45.2207} {\bibfield
  {journal} {\bibinfo  {journal} {Physical Review B}\ }\textbf {\bibinfo
  {volume} {45}},\ \bibinfo {pages} {2207} (\bibinfo {year}
  {1992})}\BibitemShut {NoStop}%
\bibitem [{\citenamefont {Fiete}(2007)}]{fiete_colloquium_2007}%
  \BibitemOpen
  \bibfield  {author} {\bibinfo {author} {\bibfnamefont {G.~A.}\ \bibnamefont
  {Fiete}},\ }\bibfield  {title} {\enquote {\bibinfo {title}
  {\textit{{Colloquium}} : {The} spin-incoherent {Luttinger} liquid},}\ }\href
  {\doibase 10.1103/RevModPhys.79.801} {\bibfield  {journal} {\bibinfo
  {journal} {Reviews of Modern Physics}\ }\textbf {\bibinfo {volume} {79}},\
  \bibinfo {pages} {801} (\bibinfo {year} {2007})}\BibitemShut {NoStop}%
\bibitem [{\citenamefont {Lai}\ and\ \citenamefont
  {Motrunich}(2010)}]{lai_two-band_2010}%
  \BibitemOpen
  \bibfield  {author} {\bibinfo {author} {\bibfnamefont {H.-H.}\ \bibnamefont
  {Lai}}\ and\ \bibinfo {author} {\bibfnamefont {O.~I.}\ \bibnamefont
  {Motrunich}},\ }\bibfield  {title} {\enquote {\bibinfo {title} {Two-band
  electronic metal and neighboring spin {Bose}-metal on a zigzag strip with
  longer-ranged repulsion},}\ }\href {\doibase 10.1103/PhysRevB.81.045105}
  {\bibfield  {journal} {\bibinfo  {journal} {Physical Review B}\ }\textbf
  {\bibinfo {volume} {81}},\ \bibinfo {pages} {045105} (\bibinfo {year}
  {2010})}\BibitemShut {NoStop}%
\bibitem [{\citenamefont {Schulz}(1990)}]{schulz_correlation_1990}%
  \BibitemOpen
  \bibfield  {author} {\bibinfo {author} {\bibfnamefont {H.~J.}\ \bibnamefont
  {Schulz}},\ }\bibfield  {title} {\enquote {\bibinfo {title} {Correlation
  exponents and the metal-insulator transition in the one-dimensional {Hubbard}
  model},}\ }\href {\doibase 10.1103/PhysRevLett.64.2831} {\bibfield  {journal}
  {\bibinfo  {journal} {Physical Review Letters}\ }\textbf {\bibinfo {volume}
  {64}},\ \bibinfo {pages} {2831} (\bibinfo {year} {1990})}\BibitemShut
  {NoStop}%
\end{thebibliography}%

\end{document}